\documentclass[aps,prd,preprint,superscriptaddress,showpacs,nofootinbib]{revtex4-1}
\usepackage{float}
\usepackage{graphicx}
\usepackage{graphics}
\usepackage{epsfig}
\usepackage{latexsym}
\usepackage{colordvi}
\usepackage{amsmath}
\usepackage{amssymb}
\usepackage{slashed}
\usepackage{xcolor}

\usepackage[title]{appendix}

\begin{document}

\preprint{\parbox[b]{1in}{ \hbox{\tt PNUTP-26/A01}  }}

\title{Revisiting the Axial Anomaly and Chiral Magnetic Effect in Dense Matter, with Applications to Axion Dark Matter}
\author{Deog Ki Hong}
\email[E-mail: ]{dkhong@pusan.ac.kr}
\affiliation{Department of Physics, Pusan National University,
             Busan 46241, Korea}
\affiliation{Extreme Physics Institute, Pusan National University, Busan 46241, Korea}        

\vspace{0.1in}

\date{\today}

\begin{abstract}
We explicitly compute the axial anomaly in dense matter and show that it takes the same form as in vacuum, even in the massless limit. This result follows from a subtle cancellation in the anomalous Ward identity between the medium-induced contributions to the divergence of the axial current and to the pseudoscalar density.
We then revisit the chiral magnetic effect in a fermionic medium coupled to an axial chemical potential in an external magnetic field. We show that, in the clean limit, the medium supports a persistent, conserved anomalous current carried by fermions. The current is determined by the axial chemical potential and suppressed by the Fermi velocity, in agreement with anomalous axial-current correlation functions.
Finally, we discuss applications to axion physics, where axion dark matter acts as an effective axial chemical potential, and estimate how helicity relaxation by impurity scattering limits the induced current in realistic conductors.
\end{abstract}

\maketitle

\newpage
\section{Introduction}
The realization of quantum anomalies associated with global symmetries in macroscopic systems has attracted considerable attention in recent years due to their ability to induce non-dissipative phenomena of topological origin. A prominent example is the axial anomaly, whose macroscopic consequences have been explored in a variety of systems, including Weyl semimetals and heavy-ion collisions, where it gives rise to anomaly-induced transport phenomena~\cite{Nielsen:1983rb,Burkov:2015hba,Armitage:2018,Kharzeev:2013ffa,Landsteiner:2016led,Landsteiner:2011cp,Kharzeev:2010gd,Basar:2013iaa}, such as the chiral magnetic effect (CME)~\cite{Vilenkin:1980fu,Fukushima:2008xe} and a large negative longitudinal magnetoresistance~\cite{Son:2012bg}.

In this article, we explicitly investigate the anomalous Ward identity (WI) associated with the axial symmetry in dense matter and revisit the CME in order to clarify the nature of the charge carriers responsible for the anomaly-induced current in an external magnetic field. We show that when fermions in dense matter are coupled to an axial chemical potential, a conserved anomalous fermion current is induced in the equilibrium ground state, in addition to the vacuum current generated by a time-dependent $\theta$ term in quantum electrodynamics (QED). This current flows along the external magnetic field and is intrinsically a medium current carried by low-energy fermionic excitations. Its origin can be traced to the helicity imbalance induced in the medium by the axial chemical potential. Since helicity is conserved at low energies for fermions in a clean medium in a magnetic field, the resulting current is persistent and does not conflict with Bloch's theorem, which forbids persistent electric currents in the ground state~\cite{Yamamoto:2015fxa}. Rather, the current-carrying state should be understood as the ground state of the Hamiltonian within a fixed-helicity sector.

The magnitude of this conserved anomalous current is determined by the axial chemical potential and is suppressed by the Fermi velocity. By contrast, we find that coupling fermions to an axial chemical potential does not induce an additional anomaly-driven equilibrium vacuum current, even when the fermions are massless.

Finally, we apply the CME to axion physics and explore its implications for the
detection of axion or axion-like particle (ALP) dark matter (DM), which naturally provides an effective axial
chemical potential for electrons even in a normal metal~\cite{Hong:2022nss}. We also estimate how helicity relaxation by impurity scattering in a realistic conductor reduces the axion-induced current when the relaxation time is short compared with the axion period.

\section{Axial anomaly and Fermi liquid}
\subsection{Fermi liquid}
It is well known that conduction electrons in a normal metal are weakly interacting quasiparticles and are accurately described by Landau's Fermi-liquid theory~\cite{Landau:1958joj,PinesNozieres}. Accordingly, their interactions with photons and phonons (or the lattice) can be treated perturbatively, and the ground state is described in terms of quasiparticles behaving as free electrons with an effective mass $m$. Because of the Pauli exclusion principle, weakly interacting fermions in a dense medium at zero temperature occupy all single-particle states up to the Fermi momentum, $p_F$, thereby forming the Fermi sea, ${\cal F}$, defined as the region of momentum space satisfying $|\vec{k}|\le p_F$. The ground state is
\begin{equation}
\left|\Omega\right>={\cal N}\prod_{\vec k\in {\cal F}}\, a^{\dagger}_{\vec k}\left|0\right>\,,	
\label{lag}
\end{equation}
where $a_{\vec k}^{\dagger}$ is the positive-energy creation operator of the fermions with momentum $\vec k$ in ${\cal F}$, and we suppress the spinor indices. The vacuum is annihilated by the positive-energy annihilation operators, $a_{\vec k}\left|0\right>=0$\,, for all $\vec k$. The constant ${\cal N}$ is fixed by the normalization condition
$\left<\Omega\right|\!\left.\Omega\right>=1$\,.	
Any change in the ground-state fermion distribution will increase its energy because of Pauli blocking. 
The number density of fermions is therefore proportional to the volume of the Fermi sea. For a homogeneous medium with spin degeneracy $g_s=2$, it is given by
\begin{equation}
n_f=\frac{p_F^3}{3\pi^2}\,.
\label{fdensity}
\end{equation}

We now consider the electron field and decompose it into plane waves,
\begin{equation}
\psi(x)=\int\frac{d^3p}{(2\pi)^3}\frac{1}{\sqrt{2E_p}}\sum_s\left(
a_{\vec p}^s u^s(p)e^{-ip\cdot x}+{b_{\vec p}^s}^{\dagger}v^s(p)e^{ip\cdot x}\right)\,
	\end{equation}
and similarly its conjugate field,
\begin{equation}
\bar\psi(x)=\int\frac{d^3p^{\prime}}{(2\pi)^3}\frac{1}{\sqrt{2E_{p^{\prime}}}}\sum_r\left(
{a_{\vec {p}^{\prime}}^r}^{\dagger} {\bar u}^r(p^{\prime})e^{ip^{\prime}\cdot x}+{b_{\vec {p}^{\prime}}^r}{\bar v}^r(p^{\prime})e^{-ip^{\prime}\cdot x}
\right)\,.
\end{equation}
Now, let us calculate the electron number density in the ground state:  
\begin{equation}
\langle n_e(\vec x)\rangle=\left<\Omega\right|\bar\psi\gamma^0\psi\left|\Omega\right>-\left<0\right|\bar\psi\gamma^0\psi\left|0\right>,	
\end{equation}
where the vacuum contribution has been subtracted to regularize the divergence.
Since the contribution from the negative-energy states cancels against the vacuum part, we are left only with the occupied positive-energy states:
\begin{equation}
\left<\Omega\right|{a_{\vec p^{\prime}}^r}^{\dagger}	{a_{\vec p}^{s}}\left|\Omega\right>= \left<\Omega\right|{a_{\vec p^{\prime}}^{r}}^{\dagger}	\,\prod_{\vec k\in {\cal F}}	\left\{{a_{\vec p}^{s}}\,,{a_{\vec k}^{s^{\prime}}}^{\dagger}\right\}	\left|0\right>=(2\pi)^32E_p\prod_{\vec k\in {\cal F}}	\delta^3(\vec p-\vec k) \left<\Omega\right|{a_{\vec p^{\prime}}^{r}}^{\dagger}\left|0\right>\,.
\end{equation}
Since the non-vanishing contribution in the momentum integration comes only from the region $(\vec p,\vec p^{\prime})\in {\cal F}$,  
one gets 
\begin{equation}
\langle n_e(\vec x)\rangle=\int_{\vec p\in {\cal F}}
\frac{d^3p}{(2\pi)^3}\frac{1}{2E_p}\left<\Omega\right|{\rm Tr}\left(\gamma^0u(p)\bar u(p)\right)\left|\Omega\right>\,.	
\label{density}
\end{equation}
We recover the electron number density, Eq.\,(\ref{fdensity}), if we take the trace over spinor indices and use the spin-sum relation~\cite{PeskinSchroeder},
\begin{equation}
\sum_su^s(p){\bar u}^s(p)=\gamma\cdot p+m\,.
\label{bilinear}
\end{equation} 

Since the fermion number is conserved in the absence of anomalies, the fermion number operator defines a superselection sector of the Hilbert space. To realize a system with a finite fermion number on average, one may couple the system to a reservoir that maintains thermal and chemical equilibrium, thereby allowing the exchange of fermions. In the grand canonical ensemble, this is implemented by introducing a chemical potential $\mu$, which acts as a Lagrange multiplier for the fermion number and fixes the average number density.
With a non-vanishing chemical potential, the energy spectrum becomes
 \begin{equation}
 E=-\mu\pm\sqrt{{\vec p}^2+m^2}\,.	
 \end{equation}
For a positive chemical potential $\mu$, particles rather than antiparticles populate the ground state, filling all single-particle states up to the Fermi surface. This forms a Fermi sea. One of the salient features of a Fermi liquid is the existence of gapless modes, corresponding to small fluctuations of fermions near the Fermi surface. If the momentum of a fermion quasiparticle is close to the Fermi momentum, $\vec p\simeq\vec p_F$, its excitation energy, measured relative to the chemical potential, becomes arbitrarily small:
\begin{equation}
E \simeq \vec v_F\cdot\left(\vec p-\vec p_F\right)\,,
\end{equation}
where $\vec v_F\equiv\vec p_F/\mu$ is the Fermi velocity.

\subsection{Axial anomaly in vacuum}
A classical symmetry often breaks down at the quantum level when the correlation functions of the associated current with other symmetry currents develop singularities at zero momentum transfer~\cite{Coleman:1982yg,Dolgov:1971ri,Frishman:1980dq}. A well-known example is the axial anomaly, which successfully accounts for the decay of the neutral pion into two photons~\cite{Bell:1969ts,Adler:1969gk}. If the axial symmetry of light quarks were non-anomalous, the decay amplitude would be further suppressed by the pion mass squared, $m_{\pi}^2$, for $m_{\pi}\to0$ and would thus be smaller by several orders of magnitude.

The axial anomaly in 3+1 dimensions arises from the triangle diagram~\cite{Bell:1969ts,Adler:1969gk} and receives no contributions beyond one loop~\cite{Adler:1969er}.  For a Dirac fermion of mass $m$ and electric charge $q$, the divergence of the axial current, $j_5^{\mu}=\bar\psi\gamma^{\mu}\gamma_5\psi$, is given by \begin{equation}
\partial_{\mu}j^{\mu}_5=2i\,m\bar\psi\gamma_5\psi+\frac{q^2}{16\pi^2}	\epsilon_{\mu\nu\rho\sigma}F^{\mu\nu}F^{\rho\sigma}\,,
\label{axial_anomaly}
\end{equation}
where the second term represents the anomaly. Before turning to the anomaly in a dense medium, we briefly review the axial anomaly in vacuum, which will serve as a useful reference for comparison.

For later convenience, we compute the axial anomaly for electrons in the Landau-level basis rather than in plane-wave states.
Electrons in a uniform magnetic field, $\vec B = B \hat z$, exhibit the Landau-level spectrum~\cite{Landau1930,JohnsonLippmann1949}
\begin{equation}
E_n(p_z)=\pm\sqrt{p_z^2+m^2+2|eB|\,n},
\end{equation}
where the Landau-level index $n$ is given by
\begin{equation}
2n = 2n_r + 1 + |m_L| - \mathrm{sign}(eB)\left(m_L + 2 s_z\right).
\end{equation}
Here, $n_r$ denotes the number of radial nodes, while $m_L$ and $s_z$ denote the orbital angular momentum and spin projections along the magnetic-field direction, respectively.

Since the eigenfunctions of Landau levels form a complete orthonormal basis, one can expand the electron field as 
\begin{equation} 
\Psi(x)=\int\!\!\sum_A \psi_A(t)U_A(\vec x)\,,
\label{mode}
\end{equation}
where the energy eigenfunctions of the Landau levels are
\begin{eqnarray} 
U_A=N_A e^{ip_z
z}e^{im_L\phi}{r_{\perp}}^{|m_L|}L_{n_r}^{|m_L|}(|eB|{r_{\perp}}^2/2)
\exp{(-|eB|{r_{\perp}}^2/4)} u_{\alpha,\beta}. 
\end{eqnarray}
Here, $N_A$ is a normalization constant, and the subscript
$A=(\alpha,\beta,n_r,m_L,p_z)$ labels the simultaneous eigenstates associated with the eigenvalues of a complete set of operators mutually commuting with the Hamiltonian.
We have defined the transverse radius $r_\perp=\sqrt{x^2+y^2}$. 
$L_n^m(x) $ is the associated
Laguerre polynomial  and
$u_{\alpha,\beta}=\chi_\alpha\otimes\eta_\beta$  is the eigenvector of
$\sigma_3\otimes\sigma_3$ where the two $\sigma_3$'s correspond to the
energy and the spin, respectively. 

We now decompose the electron field as
\begin{equation}
\Psi=\psi+\Psi_{n\ne0},
\end{equation}
where $\psi$ contains only the lowest Landau-level (LLL, $n=0$) component, while $\Psi_{n\ne0}$ contains all higher Landau-level modes.
We note that the axial anomaly arises solely from LLL electrons, whose spins are aligned antiparallel to the magnetic field. In contrast, electrons in higher Landau levels ($n \neq 0$) are spin-degenerate for a given background electromagnetic field, so their contributions to the axial charge cancel and do not contribute to the anomaly.

The LLL electrons move strictly one-dimensionally along the magnetic field. In 1+1 dimensions, the anomalous two-point function is related to the two-point correlator of vector currents through the identity $\gamma_{\mu}\gamma_5=-\epsilon_{\mu\nu}\gamma^{\nu}$: 
\begin{equation}
\Gamma_{\mu\nu}^{5}(q)=\int{\rm d}^2x\,e^{iq\cdot x}\left<j_{\mu}^5(x)j_{\nu}(0)\right>	=-\epsilon_{\mu\alpha}\Gamma^{\alpha}_{\,\,\nu}(q)\,.
\end{equation}
At one loop, the two-point correlator of vector currents in vacuum is
\begin{equation}
\Gamma^{\rm vac}_{\mu\nu}(q)=-\int\frac{{\rm d}^2p}{(2\pi)^2}\,{\rm Tr}\left(\gamma_{\mu}\frac{i}{\slashed{p}+\slashed{q}-m+i\epsilon}\gamma_{\nu}\frac{i}{\slashed{p}-m+i\epsilon}\right)\,.	
\label{vac}
\end{equation}
Since the integral is ill-defined, we evaluate it using dimensional regularization and obtain\,\footnote{The result is independent of the regularization scheme~\cite{Harvey:2005it}. A derivation using Pauli--Villars regularization can be found, for example, in Ref.~\cite{Kaplan:2009yg}.}
\begin{equation}
\Gamma^{5,{\rm vac}}_{\mu\nu}(q)=\frac{i}{\pi}\epsilon_{\mu}^{\,\,\alpha}\left(q^2g_{\alpha\nu}-q_{\alpha}q_{\nu}\right)	H(q^2,m^2)
\label{axial_divergence}
\end{equation}
with 
\begin{equation}
H(q^2,m^2)=-\int_0^1{\rm d}x\left[\frac{x(1-x)}{m^2-x(1-x)q^2-i\epsilon	}\right]\,.
\end{equation}
We see that the anomalous two-point function develops a pole at $q^2=0$ in the massless limit, $m\to0$. Even for $m\neq 0$, the anomalous WI remains valid, as the anomaly is independent of mass. To see this, we calculate the two-point function of the pseudoscalar density and the vector current at one loop, defined as 
\begin{equation}
\Gamma^{\rm vac}_{5\nu}(q)=\int{\rm d}^2xe^{iq\cdot x}\left<\bar\psi\gamma_5\psi(x)\bar\psi\gamma_{\nu}\psi(0)\right>\,.
\end{equation}
Using dimensional regularization, we obtain 
\begin{equation}
\Gamma_{5\nu}^{\rm vac}(q)=\frac{i}{2\pi}\epsilon_{\alpha\nu}q^{\alpha}\int_0^1{\rm d}x\left[\frac{m}{m^2-x(1-x)q^2-i\epsilon}\right]\,.	
\label{mass-vector}
\end{equation}
Combining the two terms, Eqs.~(\ref{axial_divergence}) and (\ref{mass-vector}), we recover the anomalous WI for the axial current in vacuum in 1+1 dimensions~\cite{Adam:1993fy},
\begin{equation}
\int{\rm d}^2x\,e^{iq\cdot x}\left<\left[\partial_{\mu}j^{\mu}_5(x)-2im\bar\psi\gamma_5\psi(x)\right]\bar\psi\gamma_{\nu}\psi(0)\right>=\frac{1}{\pi}\epsilon_{\mu\nu}q^{\mu}\,.
\label{anomal-WI}
\end{equation}
The 3+1-dimensional result in Eq.~(\ref{axial_anomaly}) is recovered by multiplying the above 1+1-dimensional expression by the Landau-level degeneracy factor $|eB|/(2\pi)$ and coupling the system to an external electromagnetic field~\cite{Kaplan:2009yg}.

%%%%

\subsection{Axial anomaly in a dense medium}
In a dense medium, gapless modes near the Fermi surface give rise to singularities in the correlation functions of symmetry currents in the limit of vanishing external momentum, $q^\mu \to 0$. The presence of such gapless modes may therefore suggest possible modifications to the anomaly structure. We show, however, that there is no additional medium contribution to the axial anomaly due to a subtle cancellation between the divergence of the axial current and the pseudoscalar density contributions in the WI. Namely, the anomalous Ward identity for the axial current in a medium of LLL electrons with chemical potential $\mu$, associated with the conserved electron number, remains exactly the same as in vacuum:
\begin{equation}
\int{\rm d}^2x\,e^{iq\cdot x}\left.\left<\left[\partial_{\mu}j^{\mu}_5(x)-2im\bar\psi\gamma_5\psi(x)\right]\bar\psi\gamma_{\nu}\psi(0)\right>\right|_{\mu\ne0}=\frac{1}{\pi}\epsilon_{\mu\nu}q^{\mu}\,.
\label{anomal-WI-medium}
\end{equation}

For the computation of the axial anomaly in a dense medium, we adopt the Landau-level basis and restore the Landau-level degeneracy factor at the end. 
At one loop, the two-point function of vector currents for 1+1-dimensional electrons in a dense medium is given by
\begin{equation}
\Gamma_{\mu\nu}(q,\mu)	=
-\int{\rm d}^2x\,e^{iq\cdot x}\,\mathrm{Tr}\left[\gamma_{\mu}S_F(x)\gamma_{\nu}S_F(-x)\right]\,,
\label{two-point}
\end{equation}
where the in-medium Feynman propagator at chemical potential $\mu$ is
\begin{equation}
S_F(x)=\int\frac{{\rm d}^2p}{(2\pi)^2}e^{-ip\cdot x}\frac{i}{(1+i\epsilon)p_0\gamma^0-\vec p\cdot \vec\gamma+\mu\gamma^0-m}\,.	\end{equation}
Integrating over $p_0$ first, we obtain 
\begin{eqnarray}
S_F(x)&=&\theta(x_0)\int\frac{{\rm d}p_z}{2\pi}\,\frac{\slashed{p}+m}{2p_0}\,\theta(p_0-\mu)e^{-ip\cdot x+i\mu x_0}	\nonumber\\
&-&\theta(-x_0)\int\frac{{\rm d}p_z}{2\pi}\,\left[\frac{\slashed{p}+m}{2p_0}\,\theta(\mu-p_0)e^{-ip\cdot x+i\mu x_0}+\frac{\slashed{p}-m}{2p_0}e^{ip\cdot x+i\mu x_0}\right]\,,
\label{med_prop}
\end{eqnarray}
where $p_0$ now denotes the positive energy $p_0\equiv\sqrt{p_z^2+m^2}>0$.

Substituting the medium propagator and rewriting the step function as $\theta(p_0-\mu)=1-\theta(\mu-p_0)$ to isolate the Fermi-sea contribution, we decompose the correlator into medium and vacuum contributions, the latter being independent of $\mu$: 
\begin{equation}
\Gamma_{\mu\nu}(q,\mu)=\Gamma_{\mu\nu}^{\rm mat}(q,\mu)+\Gamma_{\mu\nu}^{\rm vac}(q)\,,	
\end{equation}
where the vacuum part is identical to the one we introduced in Eq.~(\ref{vac}). 
After performing the $x$ integration in Eq.~(\ref{two-point}), the matter part becomes~\cite{Hong:1998tn,Hong:1999ru,Manuel:1995td}
\begin{eqnarray}
\Gamma_{\mu\nu}^{\rm mat}(q)&=&i\int\frac{{\rm d}p_z}{2\pi}\frac{1}{2p_0}\frac{1}{2k_0}\left[
\frac{\theta(p_0-\mu)\theta(\mu-k_0)}{q_0-p_0+k_0+i\epsilon}-\frac{\theta(\mu-p_0)\theta(k_0-\mu)}{q_0-p_0+k_0-i\epsilon}\right]T_{\mu\nu}^{+}(p,k)\nonumber\\
&-&i\int\frac{{\rm d}p_z}{2\pi}	\frac{1}{2p_0}
\left[\frac{\theta(\mu-p_0)}{q_0-p_0-k_0^{\prime}}\,\frac{1}{2k_0^{\prime}}\,T_{\mu\nu}^{-}(p,k^{\prime})-\frac{\theta(\mu-k^{\prime\prime}_0)}{p_0+q_0+k_0^{\prime\prime}}\,\frac{1}{2k_0^{\prime\prime}}\,T_{\mu\nu}^{-}(p,k^{\prime\prime})\right],
\label{mat_2point}
\end{eqnarray}
where $p_{\mu}=(p_0,p_z)$, $k_{\mu}=(k_0, -q_z+p_z)$, $k_{\mu}^{\prime}=(k^{\prime}_0, q_z-p_z)$, and 
$k_{\mu}^{\prime\prime}=(k_0^{\prime\prime}, -q_z-p_z)$ are two-momenta in 1+1 dimensions with positive energy, on-shell in vacuum, namely 
$p^2=k^2={k^{\prime}}^2={k^{\prime\prime}}^2=m^2$. Taking the trace over gamma matrices, one obtains for  
$l=k,k^{\prime},k^{\prime\prime}$
\begin{equation}
T_{\mu\nu}^{\pm}(p,l)=2\left(p_{\mu}l_{\nu}-g_{\mu\nu}p\cdot l+l_{\mu}p_{\nu}\right)\pm 2m^2g_{\mu\nu}\,.	
\end{equation}
In the hard-dense-loop limit, $q/\mu \to 0$, the matter contribution
can be decomposed into terms arising from modes near the Fermi surface,
$\Gamma_{\mu\nu}^{(1)}$, and contributions from the entire Fermi sea,
$\Gamma_{\mu\nu}^{(2)}$, defined by
\[
\Gamma_{\mu\nu}^{(2)}=\Gamma_{\mu\nu}^{\rm mat}-\Gamma_{\mu\nu}^{(1)}\,.
\] 
The Fermi-surface contribution, given by the first two terms in Eq.~(\ref{mat_2point}), becomes, for $q^z>0$, 
\begin{equation}
\Gamma^{(1)}_{\mu\nu}=\frac{i}{2\pi}\left(
\frac{q^z{ V}_{\mu}{ V}_{\nu}}{q_0-v_Fq^z-i\epsilon}
-\frac{q^z{\bar V}_{\mu}{\bar V}_{\nu}}{q_0+v_Fq^z+i\epsilon}
\right)	\,,
\end{equation}
where $v_F$ is the magnitude of the Fermi velocity, ${\bar V}_{\mu}=(1,v_F)$ and ${V}_{\mu}=(1,-v_F)$\,. 
Similarly, the Fermi-sea contribution is given in the hard-dense-loop limit as 
\begin{equation}
\Gamma_{\mu\nu}^{(2)}=\frac{i}{2\pi}\int_{-p_F}^{p_F}\frac{{\rm d}p_z}{\sqrt{p_z^2+m^2}}
\left[\frac{p_{\mu}{\bar p}_{\nu}+{\bar p}_{\mu}p_{\nu}}{2(p_z^2+m^2)}-g_{\mu\nu}\right]=-\frac{i}{\pi}v_F\left(g_{\mu\nu}-n_{\mu}n_{\nu}\right)\,,	
\end{equation}
where ${\bar p}_{\mu}=(p_0, -p_z)$ and $n^{\mu}$ is the unit vector normal to the medium. One readily verifies that the matter contribution to the vector
two-point function is symmetric in $\mu,\nu$ and transverse,
$q^{\mu}\Gamma^{\rm mat}_{\mu\nu}=q^{\mu}\Gamma^{\rm mat}_{\nu\mu}=0$, thereby preserving gauge
invariance. 
This transversality follows from a cancellation between the contributions from modes near the Fermi surface and those from states in the Fermi sea. In the rest frame of the medium\,\footnote{Throughout this work, the calculations are performed in the rest frame of the medium.}, $n^{\mu}=(1,0)$, one finds
\begin{equation}
q^{\mu}\Gamma^{(1)}_{\mu\nu}=\frac{i}{\pi}(0, -v_Fq^z),\quad q^{\mu}\Gamma^{(2)}_{\mu\nu}=\frac{i}{\pi}(0, v_Fq^z)\,.	
\end{equation}

\begin{figure}[t]
\vskip 0.2in
\centering 
\includegraphics[scale=0.35]{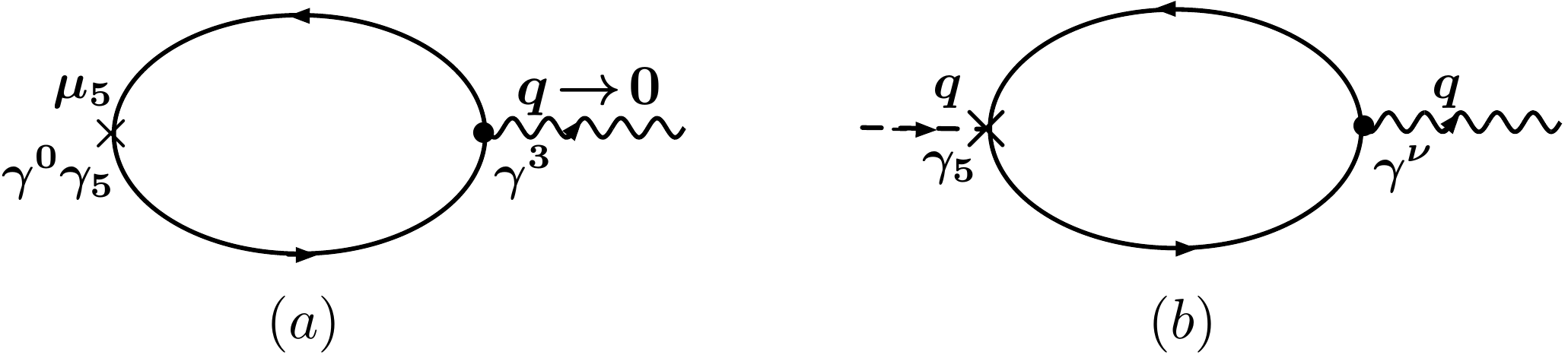}\caption{(a) The anomalous current induced by fermions coupled to an axial chemical potential. 
(b) The anomalous two-point function between the pseudoscalar density and the vector current. }\label{anomaly_cme}
\end{figure}

Similarly, the anomalous two-point function in the medium can be decomposed into a matter contribution and a vacuum contribution:
\begin{equation}
\Gamma^{5}_{\mu\nu}(q,\mu)
=
\Gamma^{5,{\rm mat}}_{\mu\nu}(q,\mu)
+
\Gamma^{5,{\rm vac}}_{\mu\nu}(q)\, .
\label{anom_two_point}
\end{equation}
Using the identity $\gamma_{\mu}\gamma_5=-\epsilon_{\mu\alpha}\gamma^{\alpha}$, the matter contribution of LLL electrons to the anomalous two-point function becomes
 \begin{equation}
 \Gamma_{\mu\nu}^{5,{\rm mat}}(q,\mu)=-\frac{i}{2\pi}\epsilon_{\mu}^{\,\,\alpha}\left[
\sum_{\vec v_F}\frac{\vec q\cdot \hat v_F{V}_{\alpha}{ V}_{\nu}}{q\cdot V -i\epsilon \vec v_F\cdot\vec q}
-2v_F
\left(g_{\alpha\nu}-n_{\alpha}n_{\nu}\right)\right],
\label{anomal_2point}
 \end{equation}
where $\hat v_F$ is the unit vector along $\vec v_F$.
We note that the term relevant for the chiral magnetic effect is
\begin{equation}
\Gamma_{03}^{5,{\rm mat}}(q,\mu)=-\frac{i}{2\pi}\left(\frac{q^z v_F^2}{q_0-v_Fq^z}-\frac{q^z v_F^2}{q_0+v_Fq^z}+2v_F\right)	\,.
\end{equation}
The gauge-invariant anomalous current of LLL electrons in dense matter, shown in Fig.~\ref{anomaly_cme}\,(a),  then becomes~\cite{Hong:2022nss}\,\footnote{As shown in Section~IV, only the matter contribution is non-vanishing.
To obtain the correct chiral magnetic current, the limits $q^{0},q^{z}\to0$ must be taken in the proper order, as the anomalous flow originates from the response of a homogeneous medium to a static, spatially uniform axial chemical potential.
 } 
\begin{equation}
\left<j^3\right>=-ie\mu_5 \lim_{q_0\to0}\lim_{q^z\to0}\left(-\Gamma^{5,{\rm mat}}_{03}(q,\mu)\frac{|eB|}{2\pi}\right)=\frac{e^2B}{2\pi^2}v_F\mu_5	\,.
\label{cme_matter}
\end{equation}
Now, let us examine whether the anomalous Ward identity, Eq.~(\ref{anomal-WI}), for the axial current receives any modifications in a medium of LLL electrons. To this end, we need to consider the two-point function of the pseudoscalar density and the vector current, shown in Fig.~\ref{anomaly_cme}\,(b):
\begin{equation}
\Gamma_{5\nu}(q,\mu)=\int{\rm d}^2xe^{iq\cdot x}\left<\bar\psi\gamma_5\psi(x)\bar\psi\gamma_{\nu}\psi(0)\right>=\Gamma_{5\nu}^{\rm mat}(q)+\Gamma_{5\nu}^{\rm vac}(q).
\end{equation}
As for the two-point function of vector currents, the medium contribution to $\Gamma_{5\nu}(q,\mu)$ at one loop is
\begin{eqnarray}
\Gamma_{5\nu}^{\rm mat}(q)&=&i\int\frac{{\rm d}p_z}{2\pi}\frac{1}{2p_0}\frac{1}{2k_0}\left[
\frac{\theta(p_0-\mu)\theta(\mu-k_0)}{q_0-p_0+k_0+i\epsilon}-\frac{\theta(\mu-p_0)\theta(k_0-\mu)}{q_0-p_0+k_0-i\epsilon}\right]N_{5\nu}^{++}(p,k)\nonumber\\
-\!\!&i&\!\!\int\frac{{\rm d}p_z}{2\pi}	\frac{1}{2p_0}
\left[\frac{\theta(\mu-p_0)}{q_0-p_0-k_0^{\prime}}\,\frac{1}{2k_0^{\prime}}\,N_{5\nu}^{+-}(p,k^{\prime})-\frac{\theta(\mu-k^{\prime\prime}_0)}{q_0+p_0+k_0^{\prime\prime}}\,\frac{1}{2k_0^{\prime\prime}}\,N_{5\nu}^{-+}(p,k^{\prime\prime})\right],
\end{eqnarray}
where $N_{5\nu}^{\sigma,\sigma^{\prime}}(p,l)\equiv{\rm Tr}\left[\gamma_5(\slashed p+\sigma m)\gamma_{\nu}(\slashed l+\sigma^{\prime} m)\right]=2m\epsilon_{\nu\rho}(\sigma l^{\rho}-\sigma^{\prime}p^{\rho})$ where $\sigma,\sigma^{\prime}$ denote the signs multiplying the mass $m$, and the two-momenta
$l=k,k^{\prime},k^{\prime\prime}$ are defined below Eq.~(\ref{mat_2point}).
In the hard-dense-loop limit, $q/\mu\to0$, we find the medium contribution
\begin{equation}
\Gamma^{\rm mat}_{5\nu}(q)=\frac{i}{2\pi}\left(\frac{mq^z}{2\mu^2}\sum_{\vec v_F}\frac{\vec q\cdot\hat v_F V_{\nu}}{q\cdot V-i\epsilon\vec v_F\cdot \vec q}+\frac{v_F}{m}\epsilon_{\nu\rho}q^{\rho}\right)%-\frac{q^z}{2m}\epsilon_{\nu\rho}\left(V^{\rho}-{\bar V}^{\rho}\right)
\,,\label{pseudoscalar}	
\end{equation}
where the first term in parentheses arises from the modes near the Fermi surface, while the second term comes from the states in the Fermi sea. We find that the medium contribution is transverse, namely $q^{\nu}\Gamma_{5\nu}^{\rm mat}=0$, thereby preserving gauge invariance. 

Combining the divergence of the axial current and the pseudoscalar contribution, we can write the medium part of the anomalous Ward identity as the sum of two terms: the Fermi-surface contribution, ${\cal A}_{\nu}^{\rm FS}$, and the Fermi-sea contribution, ${\cal A}_{\nu}^{\rm bulk}$:
\begin{equation}
{\cal A}_{\nu}^{\rm mat}(q)\equiv -iq^{\mu}\Gamma^{5,{\rm mat}}_{\mu\nu}(q)-2im\Gamma_{5\nu}^{\rm mat}(q)={\cal A}_{\nu}^{\rm FS}+{\cal A}_{\nu}^{\rm bulk}\,,	
\end{equation}
where, using $1-m^2/
\mu^2=v_F^2$,  
\begin{equation}
{\cal A}_{\nu}^{\rm FS}(q)=-\frac{1}{2\pi}
\sum_{\vec v_F}\left[\frac{\vec q\cdot\hat v_FV^{\alpha}\epsilon_{\mu\alpha}q^{\mu}V_{\nu}}{q\cdot V-i\epsilon \vec v_F\cdot\vec q}
-\frac{m^2q^z}{\mu^2}\frac{\vec q\cdot \hat v_F V_{\nu}}{q\cdot V-i\epsilon \vec v_F\cdot\vec q}\right]
=\frac{v_F}{2\pi}q^z\left(V_{\nu}+{\bar V}_{\nu}\right)
\end{equation}
and 
\begin{equation}
{\cal A}_{\nu}^{\rm bulk}(q)=\frac{1}{\pi}v_Fq^{\mu}\epsilon_{\mu}^{\,\,\rho}\left(g_{\rho\nu}-n_{\rho}n_{\nu}\right)
+\frac{1}{\pi}v_F\epsilon_{\nu\rho}q^{\rho}
\,.
\end{equation}
Combining the two contributions, we find that the anomaly contribution from the modes near the Fermi surface is exactly canceled by that from the states in the Fermi sea\,\footnote{This confirms the standard expectation that the anomaly is independent of infrared parameters of the theory, such as the chemical potential. Earlier attempts to demonstrate the density independence of the axial anomaly can be found in Refs.~\cite{Hsu:2000by,Gavai:2009vb}. }, so that
\begin{equation}
{\cal A}_{\nu}^{\rm mat}(q)=0
\,.
\label{medium_anomaly}
\end{equation}

Before concluding this section, we make a few remarks on the axial anomaly in a medium. In vacuum, the two-point function of the pseudoscalar density and the vector current is infrared finite and vanishes in the massless limit, $m\to0$, because the external momentum acts as an infrared regulator. Consequently, the pseudoscalar term does not contribute to the anomalous WI for massless fermions; see Eq.~(\ref{mass-vector}). In a medium, however, the pseudoscalar contribution survives even in the
massless limit and precisely cancels the contribution from the divergence
of the axial current, leaving no additional medium contribution to the
anomalous Ward identity.\footnote{This cancellation is consistent with
the Berry-curvature viewpoint~\cite{Son:2012wh}: although each chiral
Fermi surface carries a nonzero Berry flux, the net Chern number of a
Dirac fermion vanishes because the right- and left-handed components
contribute with opposite signs.} As a result, the anomalous WI remains identical to its vacuum form. In the massless limit, however, the pseudoscalar contribution is formally ill-defined, and a proper infrared regularization is required to obtain a consistent result.

\section{Axial chemical potential in Fermi liquid}
Recently, parity-violating Fermi liquids, whose low-energy quasiparticle dynamics near the Fermi surface breaks spatial inversion symmetry, have attracted significant interest. Parity breaking may arise from interactions with external magnetic fields, from intrinsic topological properties of the material as in Weyl semimetals~\cite{Wan2011Weyl,Son:2012bg}, or from parity-violating weak interactions~\cite{Hong:2020dwz}. In this work, we focus on parity violation induced by an external magnetic field. 

At low energies, the helicity of LLL electrons in a dense medium is conserved to a very good approximation because of Pauli blocking. Flipping the helicity requires either exciting an LLL electron to a higher Landau level, which costs an energy of order the Landau gap, or scattering it into an available state of opposite helicity in the LLL, which requires a momentum transfer of order $2p_F$. Such processes are therefore highly suppressed in the low-energy theory. (See Appendix~\ref{app:B} for more details.) 

Having a net helicity necessarily implies a net flow of LLL electrons. 
The Lagrange multiplier that enforces a helicity imbalance turns out to be the axial chemical potential. 
The Lagrangian density describing a parity-violating Fermi liquid of
LLL electrons is then
\begin{equation}
\mathcal{L}
=
\bar\psi\left(
i\slashed{\partial}-m+\mu\gamma^0+\mu_5\gamma^0\gamma^5
\right)\psi
+\mathcal{L}_{\rm int}\,,
\label{lll}
\end{equation}
where $\psi$ denotes the LLL electron field, while $\mathcal{L}_{\rm int}$
contains interactions with photons and phonons, as well as operators generated
by integrating out higher-Landau-level electrons~\cite{Hong:1997uw}.
We stress that $\mu_5$ is not a chemical potential for an exactly conserved microscopic axial charge, which is anomalous in QED. Rather, it is a Lagrange multiplier for the helicity imbalance of LLL electrons, $\Delta h$, which is an emergent and approximately conserved quantity of the low-energy effective theory below the Landau gap and the Fermi energy. The physical mechanism that prepares and maintains such a helicity imbalance is described in Appendix~\ref{app:A}. The quality of this approximate conservation, the processes that violate it, the time scale over which it holds in a realistic metal, and the resulting domain of validity of our analysis are discussed in Appendices~\ref{app:B} and~\ref{app:C}.

We first note that the LLL Lagrangian density in Eq.~(\ref{lll}) is invariant
under $U(1)_I\times U(1)_{S_z}$, where $I$ is the electron number operator and
$S_z=\frac{i}{2}\gamma^1\gamma^2$ is the spin operator along the $z$ axis
associated with the residual Lorentz symmetry. 
The (conserved) spin current is nothing but the dual of the axial current: 
\begin{equation}
j^{\mu}_s=\bar\psi\gamma^{\mu}i\gamma^1\gamma^2\psi=-\epsilon^{\mu\nu}\bar\psi\gamma_{\nu}\gamma_5\psi\,.
\end{equation}
In a dense medium, however, the low-energy theory has an enlarged symmetry. At energies below the (vector) chemical
potential and also below the Landau gap, the relevant modes are localized near the two Fermi points. Since interactions do not mix the modes at different patches
at low energy, the electron numbers are separately conserved in each patch. The symmetry of LLL electrons in a dense medium then becomes  
\begin{equation}
G=U(1)_{I_+}\times U(1)_{I_-}\times U(1)_{S_z}\,,
\end{equation}
where $I_+$ and $I_-$ are the electron number operators associated with the
positive- and negative-momentum Fermi points, respectively.   
Since LLL electrons are spin-polarized and propagate only along the magnetic-field direction, 
the number difference between the two sectors is equivalent to twice the net helicity,
\begin{equation}
\Delta h
\equiv
\frac12\left\langle\bar\psi_- \gamma_0\psi_-\right\rangle
-
\frac12\left\langle\bar\psi_+ \gamma_0\psi_+\right\rangle\,,
\end{equation}
where $\psi_+$ and $\psi_-$ correspond to LLL electrons with positive and negative momentum, respectively.

Differentiating the grand partition function with respect to the axial chemical
potential $\mu_5$,  we obtain for a system at temperature $T$ in a volume $V$ 
\begin{equation}
\frac{T}{V}\frac{\partial}{\partial \mu_5}\ln {\mathcal Z}
=
\Big\langle \bar\psi\gamma^0\gamma_5\psi\Big\rangle
=
-\Big\langle \bar\psi\gamma^3\psi\Big\rangle \, .
\label{eq:dZdmu5}
\end{equation}
In the second equality, we used the fact that the LLL electrons are
spin-polarized opposite to the magnetic field $\vec B=B\hat z$, so that
\begin{equation}
\psi=P_-\,\psi,
\qquad
P_-=\frac{1+i\gamma^1\gamma^2\,\mathrm{sign}(eB)}{2}\,,
\label{eq:Pminus}
\end{equation}
together with the projector identity, for $eB<0$,
\begin{equation}
\gamma^0\gamma^5 P_- = -\,\gamma^3 P_- \, .
\label{eq:identity}
\end{equation}
Eq.~(\ref{eq:identity}) shows that the axial chemical potential $\mu_5$ shifts the
momentum along the spin direction. Consequently, it generates a helicity imbalance between positive- and negative-helicity LLL electrons. This imbalance is conserved in the LLL medium, rather than anomalous, because the medium induces no additional contribution to the anomaly, as shown in Eq.~(\ref{medium_anomaly}). Moreover, the pseudoscalar density term does not violate helicity:
\begin{equation}
\frac{{\rm d}}{{\rm d}t}\int\!{\rm d}z\,\left<\Omega\right|\bar\psi\gamma_0\gamma^5\psi\left|\Omega\right>=2im\int\!{\rm d}z\int_{\vec p\in {\cal F}}\frac{d^3p}{(2\pi)^3}\frac{1}{2E_p}\left<\Omega\right|{\rm Tr}\left(\gamma^5u(p)\bar u(p)\right)\left|\Omega\right>=0\,.		
\end{equation}

%%%%

\begin{figure}[H]
\centering 
\includegraphics[scale=0.34]{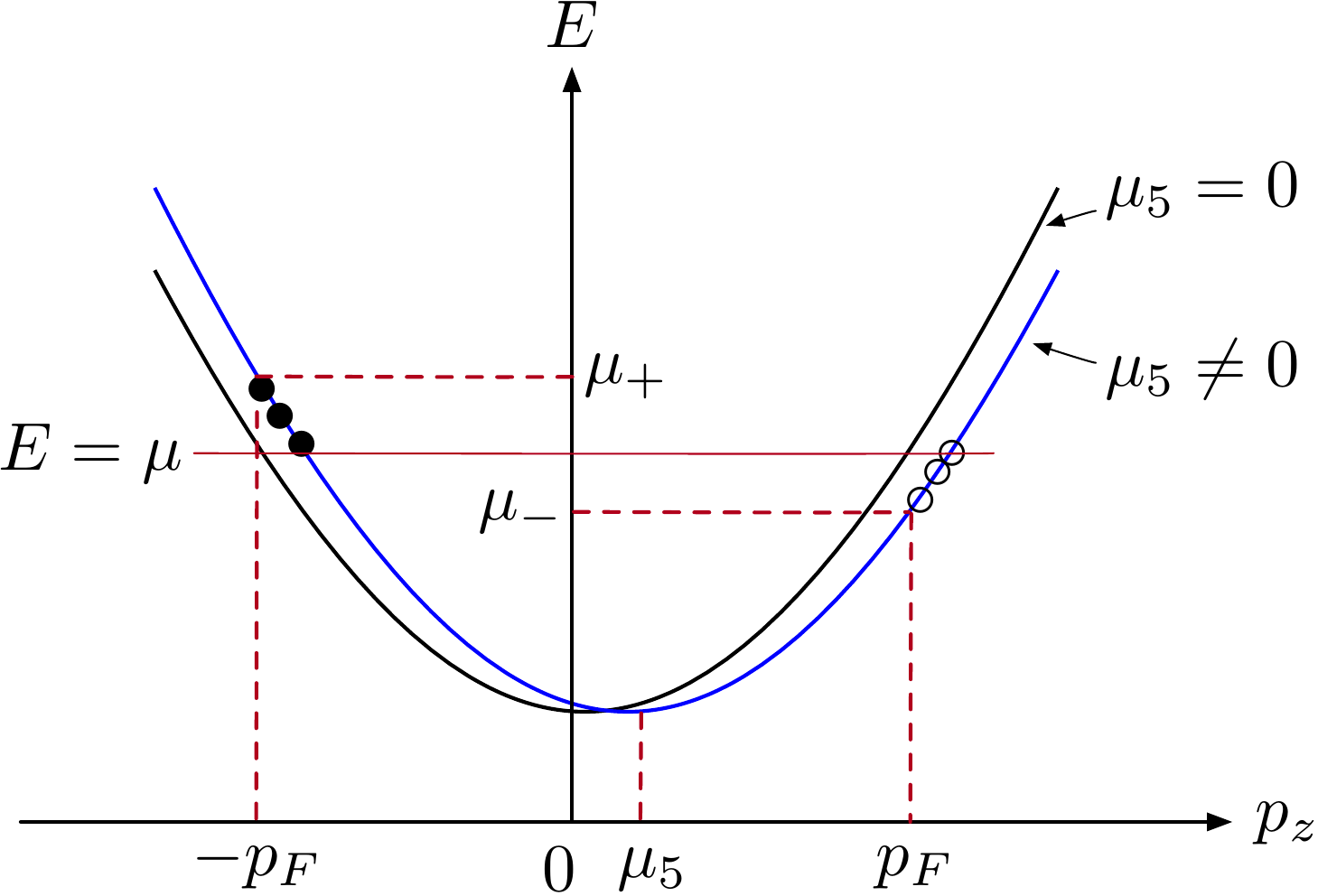}
 \caption{The axial chemical potential $\mu_5$ creates a helicity imbalance in the LLL electron medium: The black curve is the electron spectrum of normal LLL electrons for $\mu_5=0$, $E=\sqrt{p_z^2+m^2}$.  The blue curve is that of LLL electrons with $\mu_5\ne0$, $E=\sqrt{(p_z-\mu_5)^2+m^2}$, and $\mu_{\pm}=\sqrt{(p_F\pm\mu_5)^2+m^2}$. Both are plotted in the $A_z=0$ gauge.  }
 \label{shifted}
\end{figure}

Introducing an axial chemical potential into the Fermi liquid is therefore tantamount to equilibrating the system with a reservoir that supplies a constant momentum shift to the electrons along their spin direction, leading to a net flow and a net helicity~\footnote{This does not contradict the no-go theorem for equilibrium CME currents
based on the Bloch theorem~\cite{Yamamoto:2015fxa}, since the present
state is the ground state in a fixed conserved-charge (helicity) sector of the
Hilbert space.}.
The amount of electron flow generated by the axial chemical potential is then 
\begin{equation}
\left<\bar\psi\gamma^3\psi\right>=\int_{\vec p\in {\cal F}}\frac{d^3p}{(2\pi)^3}\frac{1}{2E_p}\left<\Omega\right|{\rm Tr}\left(\gamma^3u(p)\bar u(p)\right)\left|\Omega\right>\,.	
\end{equation}
Taking the trace,
\begin{equation}
{\rm Tr}\left(\gamma^3u(p)\bar u(p)\right)={\rm Tr}\left[\gamma^3\left(\gamma\cdot p+\mu_5\gamma^0\gamma_5+m\right)P_-
\right]=-2(p_z+\mu_5\,{\rm sign}(eB))\,.
\end{equation}
We see that $\mu_5$ shifts the electron momentum along the spin direction and consequently modifies the energy spectrum of the LLL electrons: in the $A_z=0$ gauge, 
\begin{equation}
E(p_z)=\sqrt{p_z^2+m^2}\longrightarrow \sqrt{(p_z-\mu_5)^2+m^2}\,.
\end{equation} 
As a result, an electric current is induced along the magnetic field direction, taken to be the $+z$ axis (see Fig.~\ref{shifted}):\begin{equation}
\left<j^3\right>\equiv-e\langle\bar\psi\gamma^3\psi\rangle=-\frac{e^2B}{4\pi^2}\int_{-p_F}^{p_F}\frac{dp_z}{E_p}(p_z-\mu_5)\,,
\label{current2}
\end{equation}
where the degeneracy factor $|eB|/(2\pi)$ arises from the integration over the perpendicular momentum, $\vec p_{\perp}=(p_x,p_y,0)$. Since the axial chemical potential shifts the LLL electron dispersion relation to
\begin{equation}
E_p=\sqrt{(p_z-\mu_5)^2+m^2}\,,
\end{equation}
it generates a helicity imbalance, proportional to $2\mu_5$, between the two Fermi points.
This imbalance is responsible for the chiral magnetic effect in dense
matter~\cite{Hong:2022nss,Hong:2025raa}: 
\begin{equation}
\left<j^3\right>=\frac{e^2B}{4\pi^2}\left[
\sqrt{(p_F+\mu_5)^2+m^2}-\sqrt{(p_F-\mu_5)^2+m^2}\right]	\simeq\frac{e^2B}{2\pi^2}\mu_5v_F\,.
\end{equation}
 The distribution (blue curve) shown in Fig.~\ref{shifted} is the minimum-energy distribution for a given helicity 
imbalance, determined by $\mu_5$, and a given electron number, determined by $\mu$. 
We note that this distribution is constrained by two conserved quantities, the helicity and the electron number.  
 When $m\ne0$, creating a helicity imbalance requires additional energy, since electrons must be removed from the states denoted by the empty circles in Fig.~\ref{shifted} and promoted to the states denoted by the filled circles in the same figure. 

\begin{figure}[t]
\vskip 0.1in
\centering 
\includegraphics[scale=0.4]{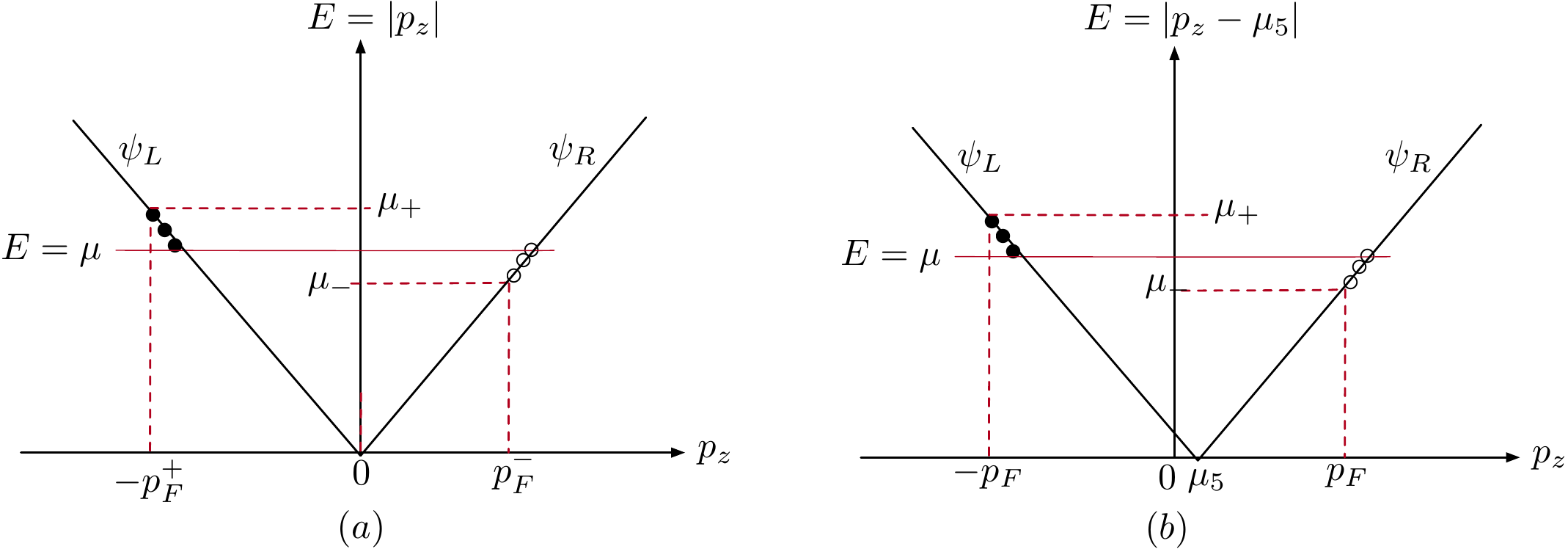}
 \caption{$(a)$ The Fermi momenta are shifted by $\mu_5$ to be $p_F^{+}=p_F+\mu_5$ and $p_F^-=p_F-\mu_5$, with $p_F=\mu$ for the massless case, while the electron momentum does not change. $(b)$ The electron momentum is shifted by $\mu_5$, while the Fermi momentum remains fixed.  Both descriptions give the same helicity imbalance and the same chiral magnetic current $\left<j^3\right>\ne0$\,. }
  \label{chiral}
\end{figure}

We also note that the axial chemical potential can be viewed either as shifting the
single-particle momentum or, equivalently, as shifting the Fermi momenta of the
two helicity sectors. Both descriptions lead to the same physical result. This
equivalence is particularly transparent in the massless limit.
Assuming the magnetic field is along $z$ direction, the medium of massless LLL electrons with fixed electron number and helicity is described by,   
 \begin{equation}
{\cal L}=\bar\psi\left(i\gamma^0\partial_t+i\gamma^3\partial_z+\mu\gamma^0+\mu_5\gamma^0\gamma_5\right)\psi\,,	
\label{lag_chiral}
\end{equation}
where we used the fact that LLL electrons move one-dimensionally along the external magnetic field, $\vec B=B\hat z$.
We decompose the LLL electron fields in the chiral basis,
\begin{equation}
\psi=\psi_L+\psi_R\quad{\rm with}\,\,\psi_L=\frac{1+\gamma_5}{2}\psi=\psi-\psi_R\,.	
\end{equation}
For all LLL electrons, the Lagrangian density Eq.~(\ref{lag_chiral}) then becomes 
\begin{equation}
{\cal L}=\bar\psi_L\left[i\gamma^0\left(\partial_t-\partial_z\right)+\mu_+\gamma^0\right]\psi_L+	\bar\psi_R\left[i\gamma^0\left(\partial_t+\partial_z\right)+\mu_-\gamma^0\right]\psi_R\,,	\end{equation}
where $\mu_\pm=\mu\pm\mu_5$\,.
We see that $\psi_L$ moves in the negative $z$ direction ($p_z<0$), carrying positive helicity, while $\psi_R$ moves in the positive $z$ direction ($p_z>0$), carrying negative helicity. The axial chemical potential creates a helicity imbalance proportional to $2\mu_5$\,[Fig.~\ref{chiral}\,$(a)$].	
Alternatively, since $\gamma^0\gamma_5\psi=-\gamma^3\psi$ for LLL electrons,  the axial chemical potential can shift the momentum, as shown in Fig.~\ref{chiral}\,$(b)$, instead of shifting the Fermi momentum, $p_F$:
\begin{equation}
p_z\to p_z-\mu_5\,.	
\end{equation}
We emphasize once again that, as one can see clearly from the massless case, the axial chemical potential shifts either the Fermi momentum, $p_F$, or the electron momentum along the spin direction. Both descriptions result in the same helicity imbalance, which is physical. 

This calculation parallels that for the vector chemical potential $\mu$, which
describes a system at temperature $T$ in a volume $V$ with a fixed number density
$n = \langle\bar\psi\gamma^0\psi\rangle$:
\begin{equation}
\frac{T}{V}\,\frac{\partial \ln\mathcal{Z}}{\partial \mu} = \langle\bar\psi\gamma^0\psi\rangle = n .
\end{equation}
The electron number density for a system with chemical potential $\mu$ in the magnetic field  is given by
\begin{equation}
\left<n_e(\mu)\right>=\frac{N_e}{V}=\left<0\right|\bar\psi\gamma^0\psi\left|0\right>-\left<0\right|\bar\psi\gamma^0\psi\left|0\right>\left.\right|_{\mu=0}\,,	
\end{equation}
where we used the fact that electron density of the vacuum  (the second term) vanishes. Explicitly,
\begin{equation}
\left<n_e(\mu)\right>=\int_0^{\mu}d\mu^{\prime}\frac{\partial}{\partial\mu^{\prime}}\left\{
(-i)\frac{|eB|}{2\pi}\int\frac{d^2p}{(2\pi)^2}\frac{{\rm Tr}\left[\gamma^0(\slashed p+\mu^{\prime}\gamma^0+m)P_-
\right]}{[(1+i\epsilon)p_0+\mu^{\prime}]^2-p_z^2-m^2}
\right\}\,.	
\label{distribution}
\end{equation}
Since the integration is finite, we shift $p_0\to p_0^{\prime}=p_0+\mu^{\prime}$ and use $1/(x+i\epsilon)=P\frac{1}{x}-\pi i\,\delta(x)\,{\rm sign}(\epsilon)$ and $({\rm sign}(z))^{\prime}=2\delta(z)$ to get, after integrating over $p_0^{\prime}$ and taking the trace, 
\begin{equation}
\left<n_e(\mu)\right>=\frac{|eB|}{2\pi}\int_0^{\mu}{d\mu^{\prime}}\int_{-\infty}^{\infty}\frac{dp_z}{2\pi}\delta(\mu^{\prime}-\sqrt{p_z^2+m^2})	
=\frac{|eB|}{2\pi}\int_{-p_F}^{p_F}\frac{dp_z}{2\pi}\,,
\label{density}
\end{equation}
where the Fermi momentum $p_F=\sqrt{\mu^2-m^2}$\,. 
The meaning of the chemical potential $\mu$ is thus clear that the ground state occupies all states up to $|p_z|=p_F\,(\equiv\sqrt{\mu^2-m^2})$.

The roles of the two chemical potentials are summarized as follows: they introduce imbalances in conserved quantities, with the magnitude of each imbalance controlled by the corresponding chemical potential. In the case of the vector chemical potential, the imbalance corresponds to a net number density of electrons, whereas in the case of the axial chemical potential, it corresponds to a net flow along the spin direction.  

The axial chemical potential enters in the same way as a constant uniform vector potential along the spin direction. Similarly, the vector chemical potential $\mu$, enters in the same way as a constant scalar potential.   If we turn on a constant gauge potential, $A_{\mu}=(\varphi, 0,0,A_z)$, the energy of LLL electrons becomes 
\begin{equation}
E=\sqrt{(p_z-eA_z-\mu_5)^2+m^2}-\mu-e\varphi\,.	
\end{equation}
Just as a constant scalar potential is not equivalent to the vector chemical potential $\mu$, a constant vector potential $eA_z$ is not equivalent to the axial chemical potential $\mu_5$. This remains true even though, in the LLL Hamiltonian, $eA_z$ and $\mu_5$ appear as the same momentum shift.

The vector chemical potential is gauge invariant. It represents a physical quantity, namely, the Fermi energy, defined as the energy difference between the vacuum and a state at the Fermi surface:
\begin{equation}
\mu
= \langle \vec p_F | H | \vec p_F \rangle
	-\langle 0 | H | 0 \rangle  \,,
\end{equation}
where $H$ is the Hamiltonian\,\footnote{Note that the energy difference is gauge invariant and a measurable physical quantity, unlike the energy itself.}.
Similarly, the axial chemical potential $\mu_5$ is a gauge-invariant physical quantity. It induces a physical shift of the kinetic momentum, $\pi_z = p_z - eA_z$, along the spin direction, according to 
\begin{equation}
\pi_z \; \to\; \pi_z - \mu_5.
\end{equation}
This shift is therefore physical and cannot be removed by a gauge transformation. It may be interpreted as a physical momentum shift measured in the rest frame of the medium. 

\section{Revisiting the Chiral Magnetic Effect }

The chiral magnetic effect is the spontaneous generation of an electric current in a dense fermionic medium subjected to an external magnetic field and in the presence of an axial chemical potential, $\mu_5$:
\begin{equation}
\vec j_{\rm cme}
=
\kappa_{\rm cme}\,\frac{e^2}{2\pi^2}\,\mu_5 \vec B\,,
\label{cme}
\end{equation}
where $\kappa_{\rm cme}=1$ in the original formula of Fukushima, Kharzeev, and
Warringa~\cite{Fukushima:2008xe}.
At first sight, the formula for the CME current is puzzling. The current flows in the absence of an external electric field and is therefore non-Ohmic. Moreover, since a magnetic field does no work on charged excitations, generating such a current in a fermionic medium appears to cost no energy. The current should therefore be persistent in the ground state of the medium.
This observation raises several basic questions. What is the physical meaning of the axial chemical potential, given that the axial charge is not exactly conserved? What are the charge carriers responsible for the CME current? Where, and how, does the dependence on the medium enter?

\begin{figure}[t]
\vskip 0.1in
\centering 
\includegraphics[scale=0.35]{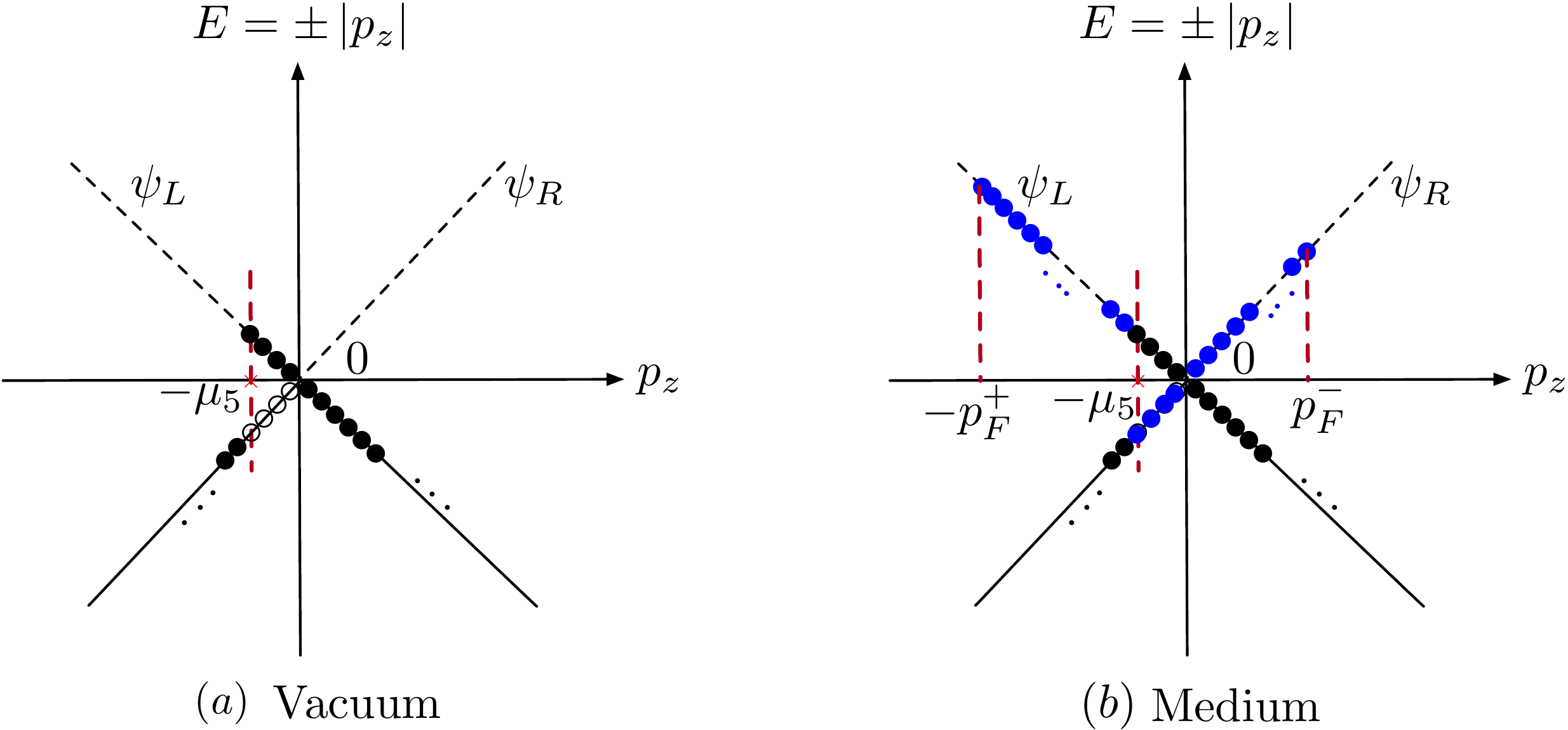}
 \caption{(a) Creation of a chirality imbalance out of the vacuum by $\mu_5$. The axial chemical potential creates left-handed fermions, denoted by bullets, and right-handed anti-fermions, denoted by holes. (b) In the medium, the holes are filled by fermions from the Fermi sea, denoted by blue bullets, leaving the net chirality imbalance unchanged. }
  \label{chiral_vacuum}
\end{figure}

Since the anomalous response function is given by the anomalous two-point function, Eq.~(\ref{anom_two_point}), the CME current has two contributions in a dense fermionic medium [Fig.~\ref{anomaly_cme}\,(a)]: 
\begin{equation}
j^{3}_{\rm cme}=j_{\rm cme}^{3,{\rm mat}}+j^{3,{\rm vac}}_{\rm cme}=ie\mu_5 \left(\Gamma^{5,{\rm mat}}_{03}+\Gamma^{5,{\rm vac}}_{03}\right)\frac{|eB|}{2\pi}\,.	
\end{equation}
Unlike the matter contribution, Eq.~(\ref{cme_matter}), the vacuum contribution is suppressed by the fermion mass, $m$, and vanishes in the static limit, $q_0\to0$:
\begin{equation}
j_{\rm cme}^{3,{\rm vac}}
=
\frac{e^2B}{2\pi^2}\mu_5\,
\frac{q_0^2}{6m^2}
\left(1+\frac{q^2}{5m^2}+\cdots\right)\, .
\end{equation}
For massless fermions, however, the vacuum contribution is nonzero if one takes the homogeneous limit, $q^z\to0$, before the static limit, $q_0\to0$, because of the anomaly pole:
\begin{equation}
j_{\rm cme}^{3,{\rm vac}}
=
ie\mu_5\lim_{q^z\to0}
\left(
\Gamma_{03}^{5,{\rm vac}}(q)\frac{|eB|}{2\pi}
\right)
=
\frac{e^2B}{2\pi^2}\mu_5\, .
\end{equation}
Another way to understand this point is to note that, for massless
fermions, the axial chemical potential creates chiral fermions out of
the vacuum, thereby producing a net chirality (helicity) imbalance of 
$2\mu_5$ (see Fig.~\ref{chiral_vacuum}\,(a))\,\footnote{For massive fermions, a shift of the momentum along the spin
direction by $\mu_5$ does not create fermions out of the vacuum and thus no CME current is generated in the vacuum.}.

In the massless limit, the matter contribution and the vacuum contribution to the anomalous two-point function are the same in the hard-dense-loop approximation, $\Gamma_{03}^{5,{\rm mat}}(q)=\Gamma_{03}^{5,{\rm vac}}(q)+{\cal O}\!\left(\frac{q_0}{\mu},\frac{q^z}{\mu}\right)$. It would be erroneous, however, to conclude that the CME current in a
medium of massless fermions is twice the vacuum contribution. Such a
conclusion neglects interactions among the fermions. Once interactions
are included, the holes in the Dirac sea created by the axial chemical
potential are removed by annihilation with fermions of opposite chirality in the Fermi sea, as shown in Fig.\,\ref{chiral_vacuum}\,(b). As a result, the net chirality, or helicity, imbalance
remains unchanged~\cite{Hong:2010hi}.
We therefore conclude that, in a static and homogeneous fermionic medium, the
CME current is given solely by the matter contribution derived in
Eq.~(\ref{cme_matter}):
\begin{equation}
j_{\rm cme}^{3}
=
v_F\frac{e^2B}{2\pi^2}\mu_5\, .
\label{our_finding}
\end{equation}
This result is in sharp contrast to the original formula,
Eq.~(\ref{cme}), in that the medium dependence now enters through the Fermi velocity, $v_F$.

The axial chemical potential that controls the helicity imbalance of fermions in a dense medium can be viewed as a thermodynamic parameter of the system, analogous to the temperature or to the fermion chemical potential $\mu$. The net helicity may arise statistically, or it may be induced by interactions with the environment. In the absence of an external magnetic field, the total helicity of massive electrons in an isotropic medium is not conserved and changes constantly through interactions, since the electrons enjoy full $SU(2)$ rotational symmetry. Once an external magnetic field is applied, however, the rotational symmetry is broken and the energy levels are quantized into Landau levels. Except for the lowest Landau level ($n=0$), all higher Landau levels are spin-degenerate, whereas LLL electrons are fully spin-polarized, so that their helicity is fixed by their direction of motion along the field. At energies below the Landau gap, LLL electrons can therefore carry a net helicity. This net helicity, parametrized by the axial chemical potential, is equivalent to a net flow of polarized fermions in the dense medium.

It has been argued that the CME is absent in strict equilibrium~\cite{Yamamoto:2015fxa,Buividovich:2013hza,Brandt:2024wlw}, in contrast with earlier lattice calculations~\cite{Yamamoto:2011ks,Yamamoto:2011gk,Muller:2016jod}. In this view, the CME has been considered to be a transient, out-of-equilibrium phenomenon originating from a time-dependent $\theta$ term in QED\,\footnote{In QED, a constant $\theta$ term has no physical significance: as a total derivative, it does not affect the equations of motion, and there are no topologically nontrivial photon vacuum configurations.}, generated, for instance, by statistical fluctuations of chiral zero modes~\cite{Kharzeev:2013ffa,Vazifeh:2013mtd,Fukushima:2012vr,Kharzeev:2016sut},
\begin{equation}
{\cal L}\ni
\theta(t)\frac{e^2}{16\pi^2}
\epsilon_{\mu\nu\rho\sigma}F^{\mu\nu}F^{\rho\sigma}\, .
\end{equation}
The time derivative of $\theta$ acts as an effective axial chemical potential that couples to the Chern--Simons density, leading to the anomalous $\theta$-dependent vacuum current
\begin{equation}
\vec j_{\rm cme}^{\,\theta,\rm vac}
=\frac{e^2}{2\pi^2}\dot\theta\,\vec B\, .
\end{equation}

The axial chemical potential $\mu_5$ appearing in Eq.~(\ref{our_finding}) is not identified with $\dot\theta$. Rather, it is a Lagrange multiplier for the approximately conserved helicity of LLL electrons in the low-energy theory, and it controls the CME current carried by LLL electrons in the medium. This current represents an equilibrium current in the ground state of a sector with nonzero net helicity. By contrast, $\dot\theta$ is the time derivative of a background field coupled to the anomalous Chern--Simons density, and the corresponding vacuum current is carried by the photon field, in analogy with a displacement current.
Thus, in principle, two distinct CME currents can arise in an electron medium subjected to a magnetic field:
\begin{equation}
\vec j_{\rm cme} =\frac{e^2}{2\pi^2}
\left(v_F\mu_5+\dot\theta\right)\vec B\, .
\end{equation}

Our analysis applies equally to quark matter, such as the quark--gluon plasma produced in heavy-ion collisions or dense matter in the cores of compact stars, provided that the quark interaction energy is smaller than the Landau gap; see, for example, Ref.~\cite{Hong:1998ka}.
 
\section{Axion Dark Matter and the Chiral Magnetic Effect}

The realization of the chiral magnetic effect requires either a time-dependent $\theta$ term in QED or an axial chemical potential coupled to fermions in a medium. Both sources violate parity. In ordinary metals, however, parity
remains unbroken up to the extremely small effects of weak interactions. Galactic dark matter, by contrast, may provide a parity-violating
background that gives rise to the CME in ordinary metals.
Indeed, axion or ALP dark matter has been shown to induce a CME current in a conductor, offering an interesting possibility for detecting axion
or ALP dark matter~\cite{Hong:2022nss,Hong:2025raa}.

The axion, a leading DM candidate, couples derivatively to electrons and also to photons through the Peccei--Quinn anomaly~\cite{Sikivie:2020zpn}.
At leading order, the interaction Lagrangian is given by 
\begin{equation}
{\cal L}_{a,{\rm int}}
=
C_{ae}\frac{\partial_{\mu}a}{f}
\bar\psi\gamma^{\mu}\gamma^5\psi+C_{a\gamma}\frac{a}{f}\frac{e^2}{16\pi^2}\epsilon_{\mu\nu\rho\sigma}F^{\mu\nu}F^{\rho\sigma}\,,
\label{axion_el}
\end{equation}
where $f$ is the axion decay constant, $C_{ae}$ is the model-dependent axion--electron coupling, which ranges from $\mathcal{O}(1)$ to values as small as $10^{-4}$ for QCD axions, depending on the underlying microscopic realization, and $C_{a\gamma}$ is the axion--photon coupling~\cite{Sikivie:2020zpn}.

Galactic axion DM is nonrelativistic and highly occupied within a de Broglie volume. It can therefore be treated as a coherent classical field
oscillating around the minimum of its potential~\cite{Sikivie:2020zpn},
\begin{equation}
a(t,\vec x)
\approx
a_0\sin\!\left(m_a t-\vec k\cdot\vec x\right)\,.
\end{equation}
Here $m_a$ is the axion mass, $v_{\rm DM}=|\vec k|/m_a\sim10^{-3}$ is the virial velocity of dark matter in our galactic halo, and the
amplitude $a_0$ is determined by the local energy density of dark matter, $\rho_{\rm DM}$. 
On laboratory length scales, the spatial variation is negligible, so one may
take
\begin{equation}
a(t)\simeq \frac{\sqrt{2\rho_{\rm DM}}}{m_a}\sin(m_a t)\,.
\end{equation}
We see immediately that the axion DM plays the role of an axial chemical potential for electrons: 
\begin{equation}
\mu_5=C_{ae}\frac{\sqrt{2\rho_{\rm DM}}}{f}\cos(m_at)\sim 2.5\times10^{-33}\,{\rm GeV}\left(\frac{\rho_{\rm DM}}{0.4\,{\rm GeV\,cm^{-3}}}\right)^{1/2}	\left(\frac{10^{12}{\rm GeV}}{f/C_{ae}}\right)\,
\label{mu5_axion}
\end{equation}
and generates an equilibrium (non-dissipative) CME current of conduction electrons coupled to axion DM, 
\begin{equation}
\vec j_{\rm cme}=v_F\,C_{ae}\frac{\sqrt{2\rho_{\rm DM}}}{f}\cos(m_at)\frac{e^2}{2\pi^2}\vec B\,.
\end{equation}
This expression holds in a clean conductor, in which helicity relaxation is slow compared with the axion oscillation, $\omega_a\tau_h\gg1$. In a realistic conductor, impurity scattering relaxes the helicity on the elastic scattering time, and the current is reduced by the factor $\omega_a\tau_h/\sqrt{1+\omega_a^2\tau_h^2}$, as shown in Appendix~\ref{app:C}.

The axial chemical potential oscillates in time with angular frequency $m_a$.
In a magnetic field, the CME current is carried entirely by the electrons in the
lowest Landau level, while the contributions of the higher Landau levels
cancel. In a good metal such as copper, many Landau levels are occupied when $|eB|\ll mE_F$ (see Fig.~\ref{fermi_points3} in Appendix~\ref{app:B}), but only the LLL is
relevant for the CME. In addition, thermal corrections to the CME current at temperature $T$ are
exponentially suppressed as $e^{-(\mu-m)/T}=e^{-E_F/T}$, where $m$ is the
electron mass and $\mu$ the chemical potential~\cite{Hong:2022nss}.

The axial chemical potential shifts the electron momentum by $\mu_5$ along the spin direction. In the presence of an external magnetic field, axion dark matter
therefore induces a helicity imbalance in the conductor after the conduction electrons equilibrate with the axion background. This helicity imbalance sources the resulting CME current.

The recent axion-search proposal LACME aims to detect precisely this
axion-induced CME current in a conductor~\cite{Hong:2022nss}. Unlike standard
axion haloscope experiments, which primarily probe the axion--photon coupling~\cite{Sikivie:1983ip},
LACME is directly sensitive to the axion--electron coupling through the CME current. Therefore, if axion DM is detected, LACME could provide a useful handle on the microscopic origin of the axion. It was also recently shown that existing axion haloscope experiments, such as ADMX and CAPP, are already sensitive to the axion--electron coupling at the
level of
\begin{equation}
g_{ae} \equiv \frac{2m C_{ae}}{f} \sim 10^{-5}
\end{equation}
in the axion-mass window~\cite{Hong:2025raa}
\begin{equation}
1\,\mu{\rm eV}\lesssim m_a\lesssim 20\,\mu{\rm eV}.
\end{equation}
The origin of this sensitivity is the additional electromagnetic radiation
inside the cavity sourced by the axion-induced CME currents of conduction
electrons on the cavity walls. We stress that the current discussed here is the microscopic source current. Its macroscopic electromagnetic signature is further affected by the electromagnetic response of the conductor; for example, the radiation emitted from the walls of a copper cavity is suppressed by $m_a^2/\sigma^2$, where $\sigma$ is the conductivity, relative to the axion--photon signal~\cite{Hong:2025raa}. These electrodynamic effects come on top of the helicity-relaxation factor derived in Appendix~\ref{app:C}, and both must be included in sensitivity estimates.

We note that the axion--electron coupling in
Eq.~(\ref{axion_el}) may be removed by a chiral rotation of the electron field. The price of this redefinition is that the axion acquires a direct coupling to the electron pseudoscalar density, while the axion--photon coupling is shifted by the vacuum axial anomaly of the electron
field~\cite{Georgi:1986df}:
\begin{equation}
-C_{ae}\frac{a}{f}\left(2im\,\bar\psi\gamma^5\psi+\frac{e^2}{16\pi^2}\epsilon_{\mu\nu\rho\sigma}F^{\mu\nu}F^{\rho\sigma}\right)\, .
\end{equation}
This form of the interaction nevertheless yields the same CME current as the derivative axion coupling, Eq.~(\ref{axion_el}).  In the limit $q^2/m^2\to0$, the vacuum contribution from the pseudoscalar density exactly cancels the anomaly contribution. The remaining matter contribution is therefore obtained from the pseudoscalar--vector correlation function in Eq.~(\ref{pseudoscalar}):
\begin{equation}
j^{3}_{\rm cme}
=
e C_{ae}\frac{a}{f}\,2im
\lim_{q\to0}
\Gamma^{\rm mat}_{5\,3}(q)\,
\frac{|eB|}{2\pi}
=
v_FC_{ae}\frac{\dot a}{f}\frac{e^2B}{2\pi^2}\,,
\end{equation}
where we have used $\dot a=-i q_0 a$ in Fourier space.  
Including the axion--photon coupling, the total current induced by axion
dark matter is therefore
\begin{equation}
\vec j=\left(v_F C_{ae}+C_{a\gamma}\right)
\frac{\dot a}{f}\frac{e^2}{2\pi^2}\vec B \, .
\end{equation}

Finally, we note that the medium-dependent CME current was derived under the assumption that the medium is clean. In practice, however, one has to consider the effect of impurities in the medium. In Appendix~\ref{app:C}, we present a simple analysis of helicity relaxation due to electron--impurity scattering, its effect on the current induced by the oscillating axion background, and the domain of validity of our analysis. 

\section{Conclusions}
The spontaneous generation of an electric current in a medium of charged
particles under an external magnetic field, known as the chiral magnetic effect, has long been regarded as a macroscopic manifestation of the quantum axial
anomaly. Revisiting the axial anomaly and the CME in dense matter, we have shown
that, besides the anomalous current induced by a time-dependent $\theta$ term---a
mechanism that has been studied extensively since its proposed relevance to
heavy-ion collisions~\cite{Kharzeev:2009pj}---there exists a medium-dependent CME
current carried by electrons, or more generally by charged fermions. This current
arises from the helicity imbalance induced by an axial chemical potential, which
shifts the fermion momentum along the spin direction.

We have explicitly shown that the axial anomaly is unaffected by the medium. The medium-dependent CME current, which is identified with the axial charge of LLL electrons in the medium by the identity in Eq.~(\ref{eq:dZdmu5}), is thus non-anomalous and persistent in equilibrium. This persistence follows from the approximate conservation of helicity in the low-energy effective theory, which holds at energies lower than the Landau gap ($E<|eB|/m$) and the Fermi energy ($E<E_F=\mu-m$). Consequently, in a medium of charged fermions with a net helicity, such as a sufficiently clean conductor placed in a sufficiently strong magnetic field, the medium-dependent CME current is a genuine equilibrium current. This finding does not conflict with the no-go theorem based on Bloch's theorem, since the current arises in the ground state of a sector with fixed helicity rather than in the unrestricted ground state of the full Hamiltonian.

The existence of this equilibrium CME current provides a microscopic understanding of the CME in dense matter. It also implies that axion or axion-like-particle dark matter, which acts as an oscillating axial chemical potential for electrons, can induce a potentially measurable CME current in conductors, provided that the helicity-relaxation time is not short compared with the axion period. This effect offers a direct probe of the axion--electron coupling and forms the basis of the LACME experiment proposed in Ref.~\cite{Hong:2022nss}.

\appendix
%\counterwithout{figure}{section}

\section{Physical mechanism of helicity imbalance}
\label{app:A}
In this appendix, we investigate a physical mechanism that establishes a sector with a fixed helicity imbalance.
Fig.~\ref{shifted} illustrates a dense system of LLL electrons with a nonzero helicity imbalance characterized by the axial chemical potential $\mu_5\ne 0$. The total electron density, on the other hand, is determined by the vector chemical potential $\mu$.

Since the LLL electrons are spin-polarized opposite to the external magnetic field, $\vec B=B\hat z$, the LLL electrons moving in the $\hat z$ direction have a negative helicity, $h=\vec p\cdot \vec S/|\vec p|=-1/2$, while the LLL electrons moving in the $-\hat z$ direction have a positive helicity $h
  =1/2$. Therefore, to create a positive net helicity, we need more electrons moving, in the rest frame of the medium, along the $-\hat z$ direction (the spin direction) than along the $+\hat z$ direction. This can be achieved if an external field coupled to electrons shifts the momentum of LLL electrons along the spin direction.      
  Suppose an external field $\varphi$ couples to the axial density of electrons,
  \begin{equation}
  {\cal L}=\bar\psi\left(i\slashed{\partial}-m\right)\psi+\varphi \bar\psi\gamma_0\gamma_5\psi\,.	
  \end{equation}
With $\vec\alpha=\gamma^0\vec\gamma$ and $\Sigma^i=\frac{i}{2}\epsilon^{ijk}\gamma^j\gamma^k=2S^i$, the chirality matrix $\gamma_5\equiv i\gamma^0\gamma^1\gamma^2\gamma^3$ satisfies the exact identity
\begin{equation}
\gamma_5=\alpha^3\Sigma^3\,,
\end{equation}
so that, for an electron with a definite spin projection along the magnetic field, $\Sigma^3\psi_s=s\psi_s$ ($s=\pm1$), the coupling becomes $\varphi\,\psi_s^{\dagger}\gamma_5\psi_s=s\varphi\,\psi_s^{\dagger}\alpha^3\psi_s$. In the Fourier space, where
\begin{equation}
\psi(t,\vec x)=\int\frac{{\rm d}^3p}{(2\pi)^3}\psi(t,\vec p)e^{i\vec p\cdot \vec x}\,,
\end{equation}
the Lagrangian density for the spin-polarized electron then becomes
\begin{equation}
{\cal L}=\bar\psi_s(t,\vec p)\left[i\gamma^0\partial_t-\gamma^3 \left(p_z-s\varphi\right)-\vec\gamma_{\perp}\cdot\vec p_{\perp}-m\right]\psi_s(t,\vec p)\,.
\end{equation}
We therefore find that the momentum of the electron along its spin is shifted by the background field $\varphi$:
\begin{equation}
\hat S\cdot\vec p\to \hat S\cdot \vec p-\varphi\,,
\end{equation}
where $\hat S\equiv2\vec S/\hbar$ denotes the unit vector along the spin direction.  For a single relativistic fermion in vacuum such a momentum shift is unphysical, as it can be removed by a unitary transformation of the field. For the dense medium of LLL electrons, however, the momentum shift is generated in the rest frame of the medium, namely the lattice in our case, and it leads to a physical observable, the net flow of LLL electrons in the rest frame, because more electrons move in one direction.

The external (background) field thus shifts the momentum of electrons along the spin direction, effectively moving negative-helicity electrons, denoted by empty circles in Fig.~\ref{shifted}, to the top of the positive-helicity distribution, denoted by filled circles. The external field therefore induces a helicity imbalance, just as an external electrostatic potential separates positive and negative charges to create a charge imbalance.

\begin{figure}[htbp]
\centering
\begin{minipage}[c]{1\linewidth}
\centering
\includegraphics[width=0.5\linewidth]{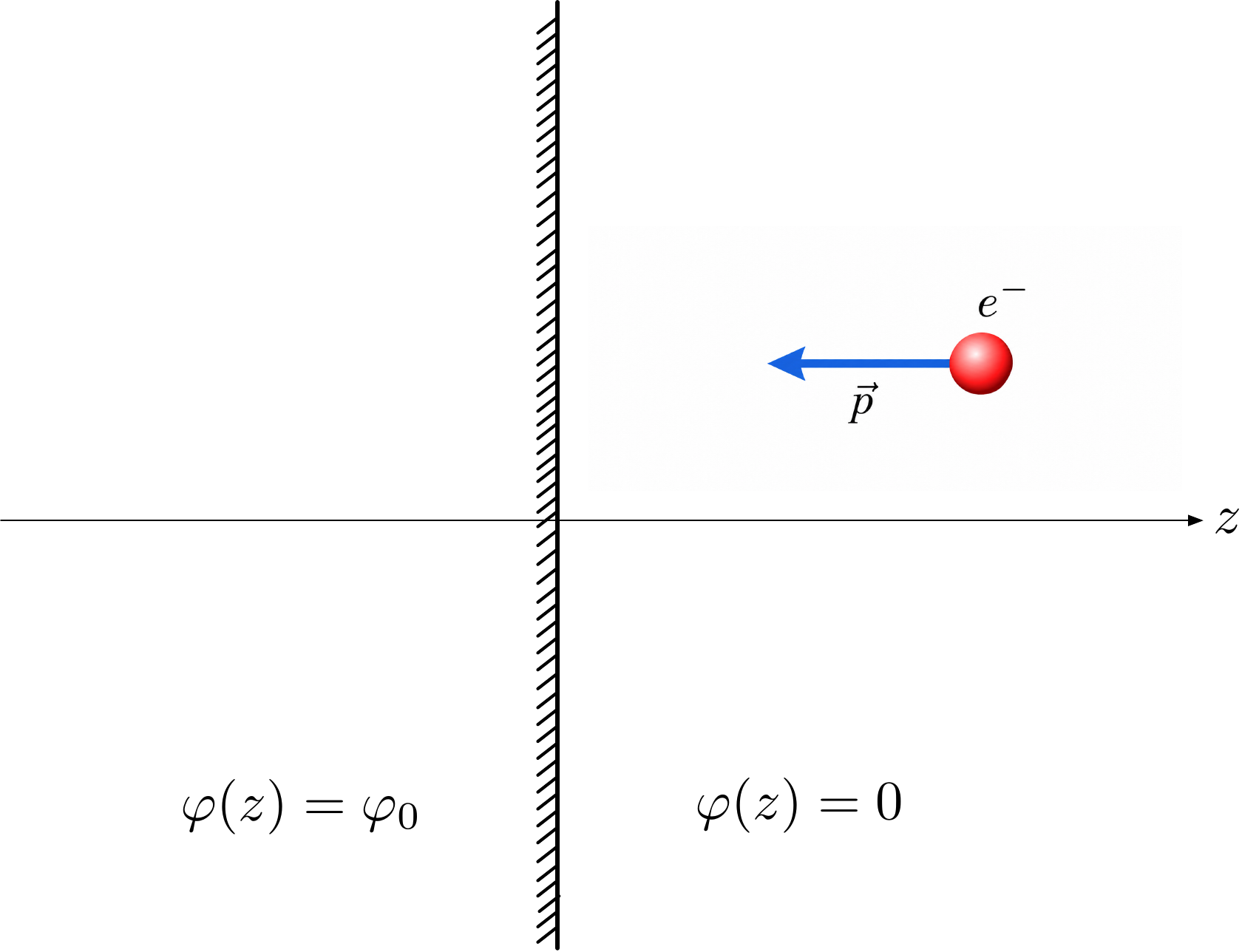}
\caption{An electron entering a wall of scalar field $\varphi$}
\label{axion_wall}
\end{minipage}
%\hfill
\end{figure}

Now, what physical mechanism generates a helicity imbalance in the dense medium of LLL electrons? Consider an electron entering a domain wall of the scalar field, $\varphi$ (see Fig.~\ref{axion_wall}): with $\varphi_0>0$
 \begin{equation}
 \varphi(z)=
 \begin{cases}
 \varphi_0, & z\le0,\\[4pt]
0, & z>0.
 \end{cases}
\end{equation}
The Hamiltonian of electrons moving along the $z$ direction (with $\beta=\gamma^0$) 
\begin{equation}
H=\alpha^3p_z+\beta m-\varphi(z)\gamma^5,
\qquad p_z=-i\partial_z.
\end{equation}
The Heisenberg equation for the electron  then gives
\begin{equation}
\frac{dp_z}{dt}=i[H,p_z]=\varphi'(z)\gamma^5=-\varphi_0\gamma_5\delta(z)
\end{equation}
Therefore the force acting on the electron becomes
\begin{equation}
F_z=-\varphi_0\left<\gamma_5\right>	\delta(z)\,.
\end{equation}
For an electron with definite spin projection, $\Sigma^3\psi_s=s\psi_s$, we obtain from $\gamma_5=\alpha^3\Sigma^3$
\begin{equation}
\left<\gamma_5\right>=s\left<\alpha^3\right>=s\,v_z\,,
\end{equation}
where we used the fact that $\alpha^3$ is the velocity operator:
\begin{equation}
\frac{dz}{dt}=i[H,z]=\alpha^3.
\end{equation}
The impulse on the electron due to the domain wall is therefore 
\begin{equation}
\Delta p_z=\int_{-\infty}^{\infty}dt\, F_z(t) =\int_{-\infty}^{\infty}\frac{dz}{v_z}F_z=s\varphi_0\,.	
\end{equation}
The same result follows from energy conservation, since inside the wall the spectrum is $E=\sqrt{(p_z-s\varphi_0)^2+m^2}$. In particular, depending on the electron spin,
\begin{equation}
\Delta p_z^{\uparrow}=+\varphi_0,\qquad
\Delta p_z^{\downarrow}=-\varphi_0.
\end{equation}
Inside the wall, the momentum of every electron is thus shifted by $\varphi_0$ along its spin. Because the wall is uniform in time, it conserves energy and does not move electrons between the two Fermi points: the occupation of each Fermi point is unchanged while the spectrum inside the wall is shifted. The medium inside the wall therefore carries the shifted distribution of Fig.~\ref{shifted} with $\varphi_0=\mu_5$, which is the lowest-energy state at fixed electron number and fixed helicity and carries a net flow. It persists as long as helicity-changing processes are absent, which we quantify in Appendix~\ref{app:B}. In the case of axion dark matter, the time derivative of the axion field plays the role of the scalar field, $\varphi=C_{ae}\dot a/f$.  Whenever electrons are immersed in the (homogeneous) axion background, they acquire a momentum shift along the spin direction, which results in a helicity imbalance in the dense medium of LLL electrons. This is similar to the way fermions acquire a mass in the background of the Higgs field in the electroweak theory.

We emphasize that the shifted distribution is the lowest-energy state only within the sector of fixed helicity. The lowest-energy state at fixed electron number alone, with a single chemical potential at both Fermi points, carries no current, in accordance with Bloch's theorem. Whether the current survives is therefore entirely a question of how fast helicity-changing processes act, compared with the time scale on which the source varies.

\section{The conservation of helicity at low energy in dense LLL electrons}  
\label{app:B}
Once a dense medium of LLL electrons develops a net helicity through a physical mechanism, the helicity imbalance may relax through interactions that change the relative populations of the two Fermi points. Possible relaxation mechanisms include transitions to higher Landau levels and backscattering of LLL electrons between opposite Fermi points, which requires a longitudinal momentum transfer of approximately $2p_F$. However, ordinary two-body electron-electron scattering within the LLL does not necessarily relax the net helicity, because energy and momentum conservation constrain the allowed changes in the right- and left-moving populations. 
\begin{figure}[htbp]
\vskip 0.1in
\centering
\begin{minipage}[c]{1\linewidth}
\centering
\includegraphics[width=0.5\linewidth]{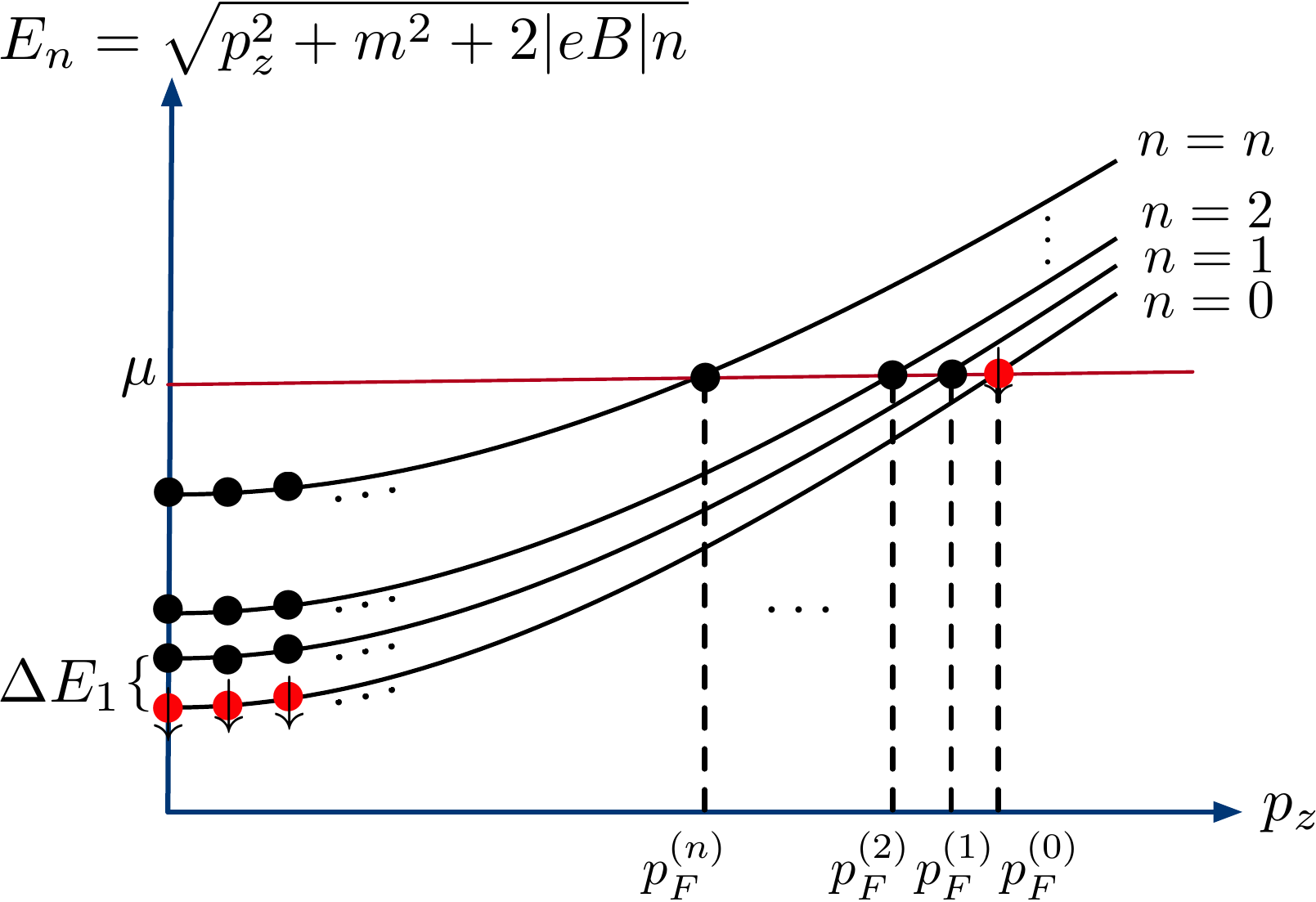}
\caption{Fermi points of electrons in the Landau levels}
\label{fermi_points3}
\end{minipage}
\end{figure}

For a typical high magnetic field in the laboratory,  $B\simeq 10\,{\rm T}$, the Landau gap is much smaller than the rest mass of electrons. The energy gap between the first excited Landau-level and the lowest Landau-level for a given momentum $p_z$ is 
\begin{equation}
\Delta E_1	=\sqrt{p_z^2+m^2+2|eB|}-\sqrt{p_z^2+m^2}\simeq \frac{|eB|}{m}\simeq 1.2\,{\rm meV}\left(\frac{B}{10\,{\rm T}}\right)\,.
\end{equation}
The Landau gap is however much larger than the typical thermal energy of electrons in cavity haloscope experiments, such as CAPP and ADMX, where the cavity temperature is maintained at a very low value, $T\lesssim 10\,{\rm mK}\, (\simeq 0.86\,\mu{\rm eV})\ll\Delta E_1$, to suppress thermal noise. On the other hand, the Fermi energy ($E_F\simeq 7\,{\rm eV}$ for copper) is larger than the Landau-level spacing. Consequently, higher Landau levels are also occupied up to their respective Fermi points, as shown in Fig.~\ref{fermi_points3}, and electrons near the Fermi points can scatter between different Landau levels, provided that the final states are unoccupied and energy and momentum are conserved. Such inter-Landau-level scattering processes can change the relative populations of the two LLL Fermi branches and thereby lead to relaxation of the net helicity carried by LLL electrons. We explicitly demonstrate, however, that these intrinsic electron-electron scattering processes are strongly suppressed at low temperatures owing to the mismatch between the Fermi momenta of different Landau levels.

For a typical metal such as copper, used in cavity haloscopes, the Fermi energy is
\begin{equation}
E_F=\mu-m\simeq7\,{\rm eV}\,,
\end{equation}
and the longitudinal Fermi momenta of the lowest and first excited Landau levels are,
respectively,
\begin{equation}
p_F^{(0)}=\sqrt{2mE_F}\simeq2.7\,{\rm keV}\,,\qquad
p_F^{(1)}=\sqrt{2mE_F-2|eB|}\simeq p_F^{(0)}-\Delta p_F\,,
\end{equation}
with $\Delta p_F\simeq|eB|/p_F^{(0)}\simeq0.22\,{\rm eV}$ for $B=10\,{\rm T}$,
corresponding to the Landau-level spacing
$v_F\Delta p_F=|eB|/m\equiv\Delta E_1\simeq1.2\,{\rm meV}$.
Since $|eB|\ll mE_F$ in a generic metal, electrons occupy many Landau levels,
$n_{\rm max}\simeq mE_F/|eB|\sim6\times10^3$ at $B=10\,{\rm T}$. LLL electrons near
the Fermi points can therefore change the net LLL helicity either by backscattering
within the LLL, which requires a momentum transfer $2p_F^{(0)}$, or by scattering
into states of higher Landau levels (and vice versa), which requires absorbing the
mismatch $\Delta p_F$. 

To study such scattering processes, let us consider the modes near the Fermi points, 
\begin{equation}
\vec p=\vec p_F^{(n)}+\vec l\,,	
\end{equation}
where the residual momentum $|\vec l|<\Lambda$. The cutoff $\Lambda$ is taken to be much smaller than the Fermi momentum difference of neighboring Landau levels, $\Lambda\ll \Delta p_F\simeq|eB|/\sqrt{2mE_F}$. (For copper, we take $\Lambda\ll 0.22\,{\rm eV}$. Note that the thermal momentum, $T/v_F\simeq 2\times10^{-4}\,{\rm eV}$ at $T=10\,{\rm mK}$, is well below this cutoff.)

\begin{figure}[htbp]
\centering
\begin{minipage}[c]{1\linewidth}
\centering
\includegraphics[width=0.7\linewidth]{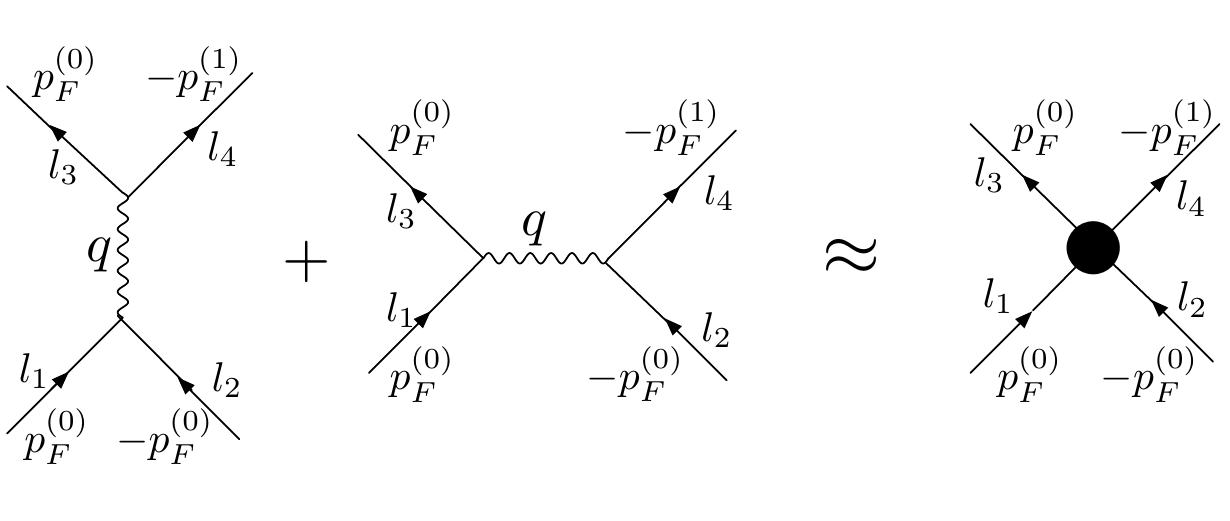}
\caption{One of the LLL electrons scatters into the $n=1$ level with Fermi momentum, $-p_F^{(1)}$. }
\label{4F}
\end{minipage}
\end{figure}

For the electrons near the Fermi points of  the $n$-th Landau level
\begin{equation}
\psi_{n}(z,t)=e^{ip_F^{(n)}z}R_{n}(z,t)+e^{-ip_F^{(n)}z}L_{n}(z,t)\,,
\end{equation}
 where $R_n$ ($L_n$) describes right (left)-moving electrons near the Fermi points at  the $n$-th Landau level.  
The effective Lagrangian for the modes near the Fermi points becomes, including the effective four-Fermi interactions, ${\cal L}_{4F}$, induced by electromagnetic interactions, 
\begin{equation}
{\cal L}_{\rm eff}=\sum_n\left[R_n^{\dagger}\left(i\partial_t+iv_F\partial_z\right)R_n+L_n^{\dagger}\left(i\partial_t-iv_F\partial_z\right)L_n\right]+{\cal L}_{4F}+{\rm h.o.}\,,	
\end{equation}
where the Fermi velocity $v_F=p^{(n)}_F/m$ is taken to be approximately the same for all Landau levels and h.o.\,\,denotes irrelevant higher-order interactions. 
Among the four-Fermi interactions induced by electromagnetic interactions, a process that does not conserve the net helicity of LLL electrons, is illustrated in Fig.~\ref{4F}. In this process, two incoming LLL electrons with opposite Fermi momenta scatter, with one of them making a transition to the first excited Landau level ($n=1$) so that the net helicity of LLL electrons changes by one unit. 
This process is kinematically forbidden at zero temperature, because the longitudinal momentum transfer is restricted to be less than the momentum cutoff of the effective theory, which is much smaller than the Fermi momentum mismatch,
\begin{equation}
\left|l_1+l_2-l_3-l_4\right|	\le 4 \Lambda\ll\Delta p_F\,.
\end{equation}
At finite temperature, however, this process for the modes near the Fermi points is possible, though highly Boltzmann suppressed. 

The photon-exchange interaction between non-relativistic electrons is dominated by
the Debye-screened Coulomb interaction, since the magnetic interaction is suppressed
by $v_F^2\simeq3\times10^{-5}$. We write the screened propagator as
\begin{equation}
V(q)=\frac{i}{q_z^2+\vec q_{\perp}^{\,2}+m_D^2}\equiv\frac{i\,G(q)}{m_D^2}\,,\qquad
G(q)=\frac{m_D^2}{q^2+m_D^2}\,,
\end{equation}
where $m_D^2\equiv2e^2mE_F/(\pi^2v_F)$ is the Debye screening mass
($m_D\simeq3.6\,{\rm keV}$ for copper).

Energy and momentum conservation with linearized dispersion require the two incoming
LLL electrons to sit at opposite Fermi points, $p_{1z}\simeq+p_{F0}$ and
$p_{2z}\simeq-p_{F0}$, with one of them scattering into the $n=1$ level near
$-p_{F1}$. Because the incoming electrons are identical, the amplitude is the
antisymmetrized sum of the direct and exchange diagrams (Fig.~\ref{4F}). The
momentum-independent (contact) part of the interaction then cancels exactly between
the two diagrams, and only $G(q)-G(0)$ contributes. In the direct channel the
momentum transfer is essentially transverse, $q_d\sim\sqrt{|eB|}\ll m_D$, giving the
negligible contribution $G(q_d)-G(0)\simeq-\vec q_\perp^{\,2}/m_D^2\sim|eB|/m_D^2
\simeq5\times10^{-5}$. In the exchange channel, however, the transfer is
$q_{e,z}\simeq2p_F\simeq1.5\,m_D$, and the backscattering component of the screened
Coulomb interaction is not small:
\begin{equation}
\kappa\equiv G(q_d)-G(q_e)\simeq1-G(2p_F)=\frac{4p_F^2}{m_D^2+4p_F^2}\simeq0.69\,.
\label{kappa_def}
\end{equation}
In the low-energy effective theory, $\kappa$ enters as the matching coefficient of
the local four-Fermi operator $\psi^\dagger_{0R}\psi^\dagger_{1L}\psi_{0L}\psi_{0R}$.
Up to $\mathcal O(1)$ differences between the direct and exchange transverse form
factors, the amplitude is
\begin{equation}
i\mathcal M_{00\to01}\simeq\frac{ie^2}{m_D^2}\,\kappa\,
\prod_{i=1}^4e^{-\vec l^{\,2}_{i\perp}/(2|eB|)}\,
\frac{l_{4x}-il_{4y}}{\sqrt{|eB|}}\,.
\label{}
\end{equation}

The transition rate per unit volume at temperature $T$ is
\begin{align}
R_{00\to01}(T)={}&\frac{1}{2!}\int\prod_{i=1}^4{\rm d}\Pi_i\,(2\pi)^4
\delta^{(4)}(p_1+p_2-p_3-p_4)\,|\mathcal M_{00\to01}|^2\nonumber\\
&\times f_0(E_1)f_0(E_2)\,[1-f_0(E_3)]\,[1-f_1(E_4)]\,,
\end{align}
where $p_i=\pm p_{F n_i}\hat z+l_i$ is measured from the Fermi point of electron $i$,
$f_n(E)=[e^{(E-\mu)/T}+1]^{-1}$ with $E-\mu\simeq\pm v_Fl_{iz}$, and
${\rm d}\Pi_i={\rm d}^3l_i/(2\pi)^3$. The factor $1/2!$ avoids double counting the
identical incoming LLL electrons. Performing the phase-space integration, we find
\begin{equation}
R_{00\to01}\sim\alpha^2\kappa^2\,\frac{|eB|^3}{m_D^4}\,\frac{T^2}{8\pi^2v_F^3}
\left[\frac{\delta/(2T)}{\sinh(\delta/(2T))}\right]^2\,,
\end{equation}
where $2\delta=v_F\Delta p_F\simeq\Delta E_1$ arises from the mismatch $\Delta p_F$
between the Fermi momenta of the lowest and first excited Landau levels.
Numerically, the prefactor is
\begin{equation}
\alpha^2\kappa^2\,\frac{|eB|^3}{m_D^4}\,\frac{T^2}{8\pi^2v_F^3}\simeq
4\times10^{11}\,{\rm cm^{-3}s^{-1}}
\left(\frac{\kappa}{0.69}\right)^{2}
\left(\frac{7\,{\rm eV}}{E_F}\right)^{2.5}
\left(\frac{B}{10\,{\rm T}}\right)^{3}
\left(\frac{T}{10\,{\rm mK}}\right)^{2}\,.
\end{equation}

Compared with helicity-preserving scattering within the LLL, for which the
Fermi-point mismatch vanishes ($\delta=0$), the longitudinal phase space is
suppressed by
\begin{equation}
\left[\frac{\delta/(2T)}{\sinh(\delta/(2T))}\right]^2\simeq
\left(\frac{\delta}{T}\right)^2e^{-\delta/T}\,,\qquad
\frac{\delta}{T}\simeq6.7\times10^2\left(\frac{B}{10\,{\rm T}}\right)
\left(\frac{10\,{\rm mK}}{T}\right)\,.
\end{equation}
The process in which one of the LLL electrons is scattered into the $n=1$ level
($00\to01$) is therefore exponentially suppressed at temperatures well below the
Landau-level spacing, $T\ll\Delta E_1$ (i.e.\ $\delta/T\gg1$), with an effective
activation energy $\delta\simeq\Delta E_1/2$. For $B=10\,{\rm T}$ and
$T=10\,{\rm mK}$ this gives $R_{00\to01}\sim10^{-275}\,{\rm cm^{-3}s^{-1}}$.
Transitions into higher levels, $n\ge2$, involve larger Fermi-momentum mismatches and
are suppressed even further. Consequently, this mechanism contributes negligibly to
the relaxation of the net helicity of LLL electrons at sufficiently low temperatures.

Other helicity-non-conserving processes involving backscattering within the LLL require a large longitudinal momentum transfer, $\Delta p_z\simeq2p_F^{(0)}$, to scatter an electron from one Fermi point to the opposite Fermi point. Electron-phonon interactions can provide such a momentum transfer. In a simplified continuum model with acoustic phonon dispersion $\omega_{\mathrm{ph}}=c_s|\mathbf q|$, the required phonon energy is
$\omega_{\mathrm{ph}}\simeq2c_sp_F^{(0)}
\simeq0.0535\,\mathrm{eV}
\simeq k_B(620\,\mathrm K)$,
where we have taken $E_F=7\,\mathrm{eV}$ and $c_s=10^{-5}$.   Phonon-induced backscattering between opposite Fermi points is therefore strongly suppressed at millikelvin temperatures; more generally, the phonon population at $T\lesssim10\,$mK is negligible, so phonon scattering is irrelevant compared with the impurity scattering discussed in Appendix~\ref{app:C}.

\section{Time scale of helicity conservation in a realistic medium}
\label{app:C}
The intrinsic (electron--electron) helicity-relaxation rate can be converted into a
relaxation time. For a small helicity imbalance $\delta n_h\simeq\chi_h\mu_5$, with
the helicity susceptibility of LLL electrons $\chi_h\simeq|eB|/(2\pi^2)$,
detailed balance gives a net relaxation rate $\delta n_h/\tau_h^{\rm intr}\simeq R_{00\to01}\,v_F\mu_5/T$ for
$\mu_5\ll T$, so that
\begin{equation}
\tau_h^{\rm intr}\simeq\frac{\chi_h T}{v_FR_{00\to01}}\sim10^{286}\,{\rm s}
\qquad (B=10\,{\rm T},\ T=10\,{\rm mK},\ E_F=7\,{\rm eV})\,,
\end{equation}
dominated by the Boltzmann factor $e^{-\delta/T}\sim10^{-292}$. The precise
prefactor is immaterial: in a clean system, the helicity of LLL electrons is
conserved for all practical purposes at millikelvin temperatures.

In a realistic metal, however, the dominant helicity-changing processes are
extrinsic. Elastic scattering off impurities and lattice defects does not
conserve momentum and can therefore supply, at no energy cost, the momentum
$2p_F^{(0)}$ needed for backscattering within the LLL, or the mismatch
$\Delta p_F$ needed for transitions into higher Landau levels at the Fermi energy.
Magnetic-impurity and spin--orbit scattering provide additional spin-flip channels
into the available states of the higher Landau levels. None of these processes is
kinematically suppressed. 
In non-magnetic metals such as copper, the spin-flip rate is expected to be smaller than that of spin-conserving impurity scattering, so it does not change the estimate below.
For an imbalance left without a source, they set the
helicity relaxation time to be of order the elastic (transport) scattering time,
\begin{equation}
\tau_h^{\rm extr}\sim\tau_{\rm imp}\simeq{\rm RRR}\times2.5\times10^{-14}\,{\rm s}\,,
\end{equation}
where RRR is the residual resistance ratio, a measure of the purity and
crystalline quality of the sample. For high-purity copper at millikelvin
temperatures, $\tau_{\rm imp}\sim10^{-12}$--$10^{-11}\,{\rm s}$. (Resolved Landau
levels require $\omega_c\tau_{\rm imp}\gg1$, with $\omega_c\simeq1.8\times10^{12}\,
{\rm s^{-1}}$ at $B=10\,{\rm T}$, which favors samples with large RRR.)

Therefore, the current-carrying state is a constrained equilibrium, in which
helicity is a conserved charge, only in a clean medium. In a real conductor,
impurity scattering relaxes the net helicity of the LLL electrons on the time
scale $\tau_h\equiv\tau_h^{\rm extr}$. Since the helicity imbalance is induced by
the axion background, $\tau_h$ is also the time for the electrons to respond to
the axion background: the axion-induced shift of the chiral energies follows $a(t)$
instantaneously, whereas the redistribution of the chiral populations that
cancels the current in full equilibrium requires helicity flips. The relevant
comparison is therefore between $\tau_h$ and the period of the axion oscillation.
The CME current of Eq.~(\ref{cme_matter}) is obtained for $\omega_a\tau_h\gg1$,
while for $\omega_a\tau_h\ll1$ it is reduced by a factor of order $\omega_a\tau_h$, as shown below.  Numerically,
\begin{equation}
\omega_a\tau_h\simeq1.5\times10^{-2}\left(\frac{m_a}{1\,\mu{\rm eV}}\right)
\left(\frac{\tau_h}{10^{-11}\,{\rm s}}\right),
\end{equation}
so the unsuppressed regime requires $m_a\gtrsim70\,\mu{\rm eV}$ for
$\tau_h\simeq10^{-11}\,{\rm s}$, whereas for lighter axions a rather clean conductor (${\rm RRR}>10^4$) is required. 

\textit{Relaxation-time treatment.}---The axion background acts on the conduction electrons as the oscillating axial chemical potential of Eq.~(\ref{mu5_axion}), with angular frequency $\omega_a=m_a/\hbar\simeq1.5\times10^{9}$--$3.0\times10^{10}\,{\rm s^{-1}}$ ($0.24$--$4.8\,$GHz) for $1\,\mu{\rm eV}\lesssim m_a\lesssim20\,\mu{\rm eV}$. A spatially uniform $\mu_5(t)$ conserves the canonical momenta, so it shifts the LLL spectrum without moving electrons between the two Fermi points. In the absence of helicity-changing processes, the electrons therefore remain in the shifted state of Fig.~\ref{shifted}, the ground state of the fixed-helicity sector, and carry the current of Eq.~(\ref{cme_matter}). Helicity-changing processes, by contrast, relax the distribution toward the equilibrium state with a single chemical potential at both Fermi points, which carries no current. (Treating the axion dark matter as a classical background is justified, since its occupation number is of order $\rho_{\rm DM}/(m_a^4v^3)\sim3\times10^{23}$ for $m_a\sim10\,\mu$eV and $v\sim10^{-3}$.)

Let $J_0(t)=v_F\,(e^2B/2\pi^2)\,\mu_5(t)$ denote the current of the fixed-helicity ground state, Eq.~(\ref{cme_matter}), and $J(t)$ the actual current. The drive changes $J$ at the rate $\dot J_0$, while helicity-changing processes relax it toward zero at the rate $1/\tau_h$:
\begin{equation}
\frac{{\rm d}J}{{\rm d}t}=\frac{{\rm d}J_0}{{\rm d}t}-\frac{J}{\tau_h}\,.
\label{relax_eq}
\end{equation}
For $\mu_5\propto\cos(\omega_at)$, the periodic steady-state solution is
\begin{equation}
J(\omega_a)=\frac{-i\omega_a\tau_h}{1-i\omega_a\tau_h}\,J_0(\omega_a)\,,
\qquad
|J|=\frac{\omega_a\tau_h}{\sqrt{1+\omega_a^2\tau_h^2}}\,|J_0|\,.
\label{relax_sol}
\end{equation}
In the collisionless regime, $\omega_a\tau_h\gg1$, the full CME current of Eq.~(\ref{cme_matter}) is recovered, in phase with $\mu_5(t)$. In the strongly relaxing regime, $\omega_a\tau_h\ll1$, the current is reduced to $|J|\simeq\omega_a\tau_h|J_0|$. Processes that conserve helicity, such as electron--electron scattering within the LLL, thermalize the electrons within each helicity sector but do not reduce $J$. For an ordinary metal there is no symmetry protecting helicity against impurity scattering, so $\tau_h\sim\tau_{\rm imp}$. For copper, $\tau_{\rm imp}\simeq{\rm RRR}\times2.5\times10^{-14}\,$s gives
\begin{equation}
\omega_a\tau_h\simeq3.8\times10^{-5}\times{\rm RRR}\left(\frac{m_a}{1\,\mu{\rm eV}}\right),
\label{omega_tau_RRR}
\end{equation}
which for ${\rm RRR}=100$ is $4\times10^{-3}$--$8\times10^{-2}$ over $1$--$20\,\mu$eV. The full current is approached only for ${\rm RRR}\gtrsim10^4$, or in materials in which helicity (chirality) relaxation is much slower than momentum relaxation.

\textit{Domain of validity.}---Our analysis does not assume that electrons occupy only the LLL: higher Landau levels are occupied, but their spin-paired states give canceling contributions, so the CME current is carried entirely by LLL electrons. The conditions for the validity of the analysis are as follows.

(i) \textit{Temperature.} The temperature must be small compared with the Landau-level spacing, $\Delta E_1=|eB|/m$:
\begin{equation}
\frac{T}{\Delta E_1}\simeq7\times10^{-4}\left(\frac{T}{10\,{\rm mK}}\right)\left(\frac{10\,{\rm T}}{B}\right)\ll1\,.
\label{cond_T}
\end{equation}

(ii) \textit{Interaction scale.} At finite temperature, the characteristic longitudinal momentum deviation from the Fermi points is set by the thermal energy, whereas the transverse momentum scale is set by the inverse magnetic length, $|l_z|\sim T/v_F$ and $|l_\perp|\sim\sqrt{|eB|}$. The ultraviolet scale of the screened Coulomb interaction is the Debye mass $m_D$, which is determined by the Fermi energy $E_F=\mu-m$. For copper at $B=10\,$T and $T=10\,$mK,
\begin{equation}
\frac{T}{v_F}\sim2\times10^{-4}\,{\rm eV}\ll\Delta p_F\sim0.2\,{\rm eV}\ll\sqrt{|eB|}\simeq24\,{\rm eV}\ll m_D\simeq3.6\,{\rm keV}\,.
\label{cond_scales}
\end{equation}
Since the momentum-independent contact part of the interaction cancels between the direct and exchange contributions, only $G(q)-G(0)$ survives. Projected onto the LLL, the forward-scattering coupling, with transfer $q\sim l_\perp$, is
\begin{equation}
g_{\rm fwd}\simeq\int\!\frac{{\rm d}^2l_\perp}{(2\pi)^2}\,\frac{4\pi\alpha}{l_\perp^2+m_D^2}\,\frac{l_\perp^2}{m_D^2}\,e^{-l_\perp^2/(2|eB|)}\simeq\frac{4\alpha|eB|^2}{m_D^4}\,,
\end{equation}
which is suppressed by $|eB|/m_D^2$. For electrons at opposite Fermi points, the exchange contribution carries $q_z\simeq2p_F$, and the backscattering coupling is not suppressed by this factor:
\begin{equation}
g_{\rm bs}\simeq\frac{4\pi\alpha\kappa}{m_D^2}\,\frac{|eB|}{2\pi}=\frac{2\alpha\kappa|eB|}{m_D^2}\,,
\end{equation}
with $\kappa\simeq0.69$ given in Eq.~(\ref{kappa_def}). Since $m_D^2/|eB|\simeq2\times10^4$ for copper at $B=10\,$T, both couplings are small, and the LLL electrons are weakly interacting, with the interaction dominated by $2p_F$ backscattering. For excitations of energy $E\lesssim T$, the interaction-induced energy shift is $\delta E\sim(g_{\rm bs}/\pi v_F)\,E\ll\Delta E_1$.

(iii) \textit{Disorder broadening.} In a real conductor, Landau levels are broadened by elastic impurity scattering, $\Gamma_{\rm imp}=\hbar/\tau_{\rm imp}$. Landau quantization, and hence our analysis, requires $\Gamma_{\rm imp}\ll\Delta E_1$, i.e., $\omega_c\tau_{\rm imp}\gg1$ with the cyclotron frequency $\omega_c=|eB|/m$. For copper,
\begin{equation}
\omega_c\tau_{\rm imp}\simeq0.04\times{\rm RRR}\left(\frac{B}{10\,{\rm T}}\right),
\label{cond_disorder}
\end{equation}
so the condition is satisfied for high-purity copper (${\rm RRR}\gtrsim100$) at millikelvin temperatures and $B\simeq10\,$T, but not for room-temperature or highly disordered conductors.

(iv) \textit{Hierarchy.} Including the condition $\hbar/\tau_h\ll\hbar\omega_a$ derived above, the conditions for the full CME current in an ordinary metal are
\begin{equation}
T\,,\ \frac{\hbar}{\tau_h}\ll\hbar\omega_a\,,\qquad
\frac{\hbar}{\tau_{\rm imp}}\ll\frac{|eB|}{m}\ll E_F\,.
\label{hierarchy}
\end{equation}
For representative haloscope parameters, $T=10\,$mK ($0.86\,\mu$eV), $B=10\,$T ($|eB|/m\simeq1.2\,$meV), $E_F=7\,$eV, and $\hbar\omega_a\simeq1$--$20\,\mu$eV, the conditions $T\ll\hbar\omega_a$ (for most of the mass window) and $|eB|/m\ll E_F$ are satisfied. For copper with ${\rm RRR}=100$, $\hbar/\tau_{\rm imp}\simeq0.26\,$meV, so Landau levels are resolved; but since $\tau_h\sim\tau_{\rm imp}$ in an ordinary metal, $\hbar/\tau_h\gg\hbar\omega_a$, and the current is in the strongly relaxing regime of Eq.~(\ref{relax_sol}).

\acknowledgments
We thank Urs Wiedemann and Seokhoon Yun for interesting comments. This work was supported by Basic Science Research Program through the National Research Foundation of Korea (NRF) funded by the Ministry of Education (NRF-2017R1D1A1B06033701).  The author used Claude (Anthropic) during the revision of this manuscript to assist with language editing and with a limited number of calculations and cross-checks. All results presented were derived or confirmed independently by the author, who takes full responsibility for the content of this work.

%\begin{appendices}

\end{document}